\documentclass[english]{article}
\usepackage[T1]{fontenc} \usepackage[latin9]{inputenc} \usepackage{geometry}
\usepackage{amsmath} \usepackage{amssymb} \usepackage{setspace}
\PassOptionsToPackage{normalem}{ulem} \usepackage{ulem} \usepackage{cite}
\usepackage{dsfont} \usepackage{mathrsfs} \usepackage{babel} \usepackage{url}
\usepackage{framed} \usepackage{color} \usepackage{needspace}
\usepackage{empheq}
\usepackage{graphicx} 

\usepackage[capbesideposition=inside,
facing=yes,capbesidesep=quad]{floatrow}

\makeatletter

\let\@fnsymbol\@arabic

\makeatother

\newcommand{\Csecond}{\widetilde{\mathbb{C}}}

\newcommand{\Cfourth}{\overline{\mathbb{C}}}
\newcommand{\Csym}{\mathbb{C}}

\newcommand{\Ls}{\widehat{\mathbb{L}}}
\newcommand{\Lsym}{\mathbb{L}}
\newcommand{\E}{\overline{\mathbb{E}}_{\mathrm{cross}}}

\newcommand{\Coe}{\Cfourth_{e}}
\newcommand{\Ch}{\Csym_{\mathrm{micro}}}

\newcommand{\mm}{\mu_{\mathrm{macro}}}
\newcommand{\lm}{\lambda_{\mathrm{macro}}}
\newcommand{\mh}{\mu_{\mathrm{micro}}}
\newcommand{\lh}{\lambda_{\mathrm{micro}}}
\newcommand{\me}{\mu_{e}}
\newcommand{\mc}{\mu_{c}}
\newcommand{\lle}{\lambda_{e}}

\newcommand{\mLc}{\mu\, L_{c}^{2}}
\newcommand{\mLd}{\mu\, L_{d}^{2}}

\newcommand{\R}{\mathbb{R}}
\newcommand{\nablau}{\,\nabla u\,}
\newcommand{\p}{\,{P}}
\newcommand{\nablap}{\nabla P}
\newcommand{\Curl}{\,\mathrm{Curl}}
\newcommand{\dev}{\, \mathrm{dev}}
\newcommand{\Div}{\mathrm{Div}}
\newcommand{\tr}{\, \mathrm{tr}}
\newcommand{\sym}{\, \mathrm{sym}}
\newcommand{\Sym}{\mathrm{Sym}}
\renewcommand{\skew}{\, \mathrm{skew}}

\newcommand{\so}{\mathfrak{so}}
\newcommand{\X}{\,}
\newcommand{\x}{\cdot}
\newcommand{\langlenew}{\,\big\langle\,}
\newcommand{\ranglenew}{\,\big\rangle}

\newcommand{\sll}{\mathfrak{sl}\left( 3\right) }

\newcommand{\soo}{\mathfrak{so}\left( 3\right) }
\newcommand{\Symtre}{\mathrm{Sym}\left( 3\right)}
\renewcommand{\skew}{\, \mathrm{skew}}

\newcommand{\devsym}{\dev\sym}

\newcommand{\id}{\mathds{1}}

\title{\vspace{-1.5cm}Complete band gaps including non-local effects occur only in the relaxed micromorphic
	model}

\author{Angela Madeo\footnote{Angela Madeo, corresponding author, angela.madeo@insa-lyon.fr, LGCIE, INSA-Lyon, Université de Lyon, 20 avenue	Albert Einstein, 69621, Villeurbanne cedex, France}\; and
	Patrizio Neff\,\footnote{Patrizio Neff, patrizio.neff@uni-due.de, Head of Chair for 	Nonlinear Analysis and Modelling, Fakultät für Mathematik, Universität Duisburg-Essen,  Mathematik-Carrée, Thea-Leymann-Straße 9, 45127 Essen}\; and
Marco Valerio 	d'Agostino\footnote{Marco Valerio d'Agostino, marco-valerio.dagostino@insa-lyon.fr, LGCIE, INSA-Lyon, Université de Lyon, 20 avenue Albert Einstein, 69621, Villeurbanne cedex, France}\; and 
	Gabriele Barbagallo\footnote{Gabriele Barbagallo, gabriele.barbagallo@insa-lyon.fr, LaMCoS-CNRS \& LGCIE, INSA-Lyon, Universitité de Lyon, 20 avenue Albert Einstein, 69621, Villeurbanne cedex, France}}

\begin{document}
	\maketitle 
	\addtocounter{footnote}{5} 

	\vspace{-0.7cm}	
	\begin{abstract}
		In this paper we substantiate the claim implicitly made in previous works that the relaxed micromorphic model is the \textbf{only linear, isotropic, reversibly elastic, nonlocal generalized continuum model} able to describe complete band-gaps on a phenomenological level. To this end, we recapitulate the response of the standard Mindlin-Eringen micromorphic model with the full micro-distortion gradient $\nablap$, the relaxed micromorphic model depending only on the $\Curl\p$ of the micro-distortion $\p$, and a variant of the standard micromorphic model in which the curvature depends only on the divergence $\Div \p$ of the micro distortion. The $\Div$-model has size-effects but the dispersion analysis for plane waves shows the incapability of that model to even produce a partial band gap. Combining the curvature to depend quadratically on $\Div\p$ and $\Curl\p$ shows that such a model is similar to the standard Mindlin-Eringen model which can eventually show only a partial band gap.	\end{abstract}
	
	\vspace{0.3cm}
	
	\hspace{-0.55cm}\textbf{Keywords}: relaxed micromorphic model, band gaps, generalized continuum models, long wavelength limit, macroscopic consistency, Cauchy continuum, homogenization, multi-scale modeling, parameter identification, non-redundant model 
	
	\vspace{0.3cm}
	
	\hspace{-0.55cm}\textbf{AMS 2010 subject classification}:  74A10 (stress), 74A30 (nonsimple materials), 74A35 (polar materials), 74A60 (micromechanical theories), 74B05 (classical linear elasticity), 74E15 (crystalline structure), 74M25 (micromechanics), 74Q15 (effective constitutive equations)
	
\vspace{-0.2 cm}
	\tableofcontents
	
	\newpage
	
	\addtocontents{toc}{\vspace{-0.5 cm}}
	
\section{Introduction}

The micromorphic model \cite{eringen1964nonlinear,eringen1966mechanics,eringen1999microcontinuum,mindlin1964micro,germain1973method} is a generalized continuum model suitable for the effective multi-scale-description of heterogeneous media with strong contrast of the mechanical properties at the microscopic level through the introduction of a \textbf{characteristic length scale}  $L_{c}$. It allows to incorporate new effects which extend the classical linear elastic description, e.g. \textbf{size-effects}, the \textbf{dispersion of waves} and the possibility of micro-motions which are in principle independent of the macro motions.  This model couples the \textbf{macroscopic displacement field} $ u:\Omega\subset\R^{3}\rightarrow\R^{3}$ and an \textbf{affine substructure deformation} attached at each macroscopic point encoded by the \textbf{micro-distortion field} $\p:\Omega\subset\R^{3}\rightarrow\R^{3\times3}$.

The curvature contribution in the micromorphic model conceptually determines how the substructure interacts with itself and the associated characteristic length is a measure of the range of action of such micro-structure related deformation modes. In this sense we call the full-gradient contribution $\rVert\nablap\lVert^2$ (or any other curvature term essentially controlling $\nablap$ ) of \textbf{strong interaction} type: neighboring substructures feel the presence of each other, or, what is the same, the generated moment stresses depend on $\nablap$.

To the contrary, in the relaxed micromorphic model, the corresponding moment stresses depend only on $\Curl\p$, therefore there is some freedom between particles but a connection of neighboring cells is still possible thanks to tangent micro-interactions. Certain substructure deformations are energetically free (in fact all compatible parts $\nabla\vartheta$ in $\p$ are not taken into account) while the model remains reversible elastic and energy-conservative. We may call this a \textbf{weak interaction}. As a matter of fact, the wording \textbf{relaxed} is motivated by this observation.

In the $\Div$-model to be introduced below, a similar effect appears. The corresponding moment stresses depend only on $\Div\p$. Therefore, substructure deformations of the type $\p=\Curl\,\zeta+\nabla\vartheta$, where $\zeta:\R^{3\times3}\rightarrow\R^{3\times3}$ is arbitrary and $\vartheta:\R^{3}\rightarrow\R^{3}$ satisfies $\Delta\vartheta\equiv0$ are energetically free. This model is, hence, also of weak-interaction type.

It is therefore intriguing that it is not simply \textbf{weak} versus \textbf{strong interaction} that determines the possibility of band gaps but there is some further hidden mechanism in the relaxed micromorphic model which, together with a positive Cosserat couple modulus $\mc>0$, is decisive for the ability to model complete band gaps and still being nonlocal.

In further contributions we will provide more detailed arguments concerning the fact that the residual freedom which is peculiar of the relaxed micromorphic model is a key feature for allowing band-gap behaviors. In fact, internal variable models (i.e. models with no dependence on the derivatives of $\p$ at all) still allow the description of complete band gaps \cite{pham2013transient,sridhar2016homogenization}, but they loose any information concerning non-locality. Non-local effects are intrinsically present in micro-structured materials, even if in some particular cases their overall effect can be, in a first approximation, neglected. Nevertheless, as far as the contrast of mechanical properties between adjacent unit cells at the micro level becomes more pronounced, non local effects are sensible to rapidly become non-negligible. In this optic, a model including non-locality is to be considered as the natural choice for modeling the mechanical behavior of metamaterials.    

This paper is now structured as follows. First, we introduce the relaxed micromorphic model with an augmented curvature energy depending also on $\Div \p$. The governing equations for wave propagation are derived and the plane wave ansatz is introduced. Then we particularize the result for specific cases and show the resulting dispersion curves for each of them. Finally, we provide for completeness the standard Mindlin-Eringen micromorphic model together with its dispersion curves thus recognizing that it is equivalent to a particular case of the augmented relaxed micromorphic model with $\Div \p$.

\section{The relaxed micromorphic continuum  with $\rVert\Curl\p\lVert^{2}$ and $\rVert\Div\p\lVert^{2}$}

The \textbf{relaxed micromorphic model} \cite{neff2015relaxed,neff2014unifying,madeo2014band,madeo2015wave} has been introduced in 2013 in  \cite{neff2014unifying} and endows the standard Mindlin-Eringen's representation
with more geometric structure by reducing the curvature energy term to depend \textbf{only} on the \textbf{second order dislocation density tensor} $\alpha=-\Curl \p$. Here, we additionally consider also a curvature term depending on $\Div\p$. The strain energy density for the resulting micromorphic continuum can be written as:

\begin{align}
	W=&\underbrace{\me\,\lVert \sym\left(\nablau-\p\right)\rVert ^{2}+\frac{\lle}{2}\left(\mathrm{tr} \left(\nablau-\p\right)\right)^{2}}_{\mathrm{{\textstyle isotropic\ elastic-energy}}}	+\hspace{-0.1cm}\underbrace{\mc\,\lVert \skew\left(\nablau-\p\right)\rVert ^{2}}_{\mathrm{\textstyle rotational\   elastic\ coupling}		}\hspace{-0.1cm} \label{eq:Ener-General}\\
	& \quad 
	+\underbrace{\mh\,\lVert \sym \p\rVert ^{2}+\frac{\lh}{2}\,\left(\mathrm{tr} \p\right)^{2}}_{\mathrm{{\textstyle micro-self-energy}}}
	+\underbrace{\frac{\mLc}{2} \,\lVert \Curl \p\rVert^2+\frac{\mLd}{2} \,\lVert \Div \p\rVert^2}_{\mathrm{\textstyle simple\ isotropic\ curvature}}\,,
	\nonumber  
\end{align}
where all the introduced elastic coefficients are assumed to be constant.
This decomposition of the strain energy density, valid in the isotropic,
linear-elastic case, has been proposed in \cite{neff2015relaxed,ghiba2014relaxed}
where well-posedness theorems have also been proved. It is clear that
this decomposition introduces a limited number of elastic parameters
and we will show how this may help in the physical interpretation
of these latter. Positive definiteness of the potential energy implies
the following simple relations on the introduced parameters
\begin{equation}
	\mu_{e}>0,\ \quad\mu_{c}\geq0,\ \quad 3\lambda_{e}+2\mu_{e}>0,\ \quad\mh>0,\ \quad 3\lh+2\mh>0,\ \quad\mLc>0,\qquad\mLd>0.\label{DefPos}
\end{equation}
We need to remark that this model variant is  not strictly positive definite in the sense of the standard Mindlin-Eringen model. One of the most interesting features of the proposed strain energy
density is the reduced number of elastic parameters which are needed
to fully describe the mechanical behavior of a micromorphic continuum.
Indeed, each parameter can be easily related to specific micro and
macro deformation modes. 

Comparing classical linear elasticity with our new relaxed model for $L_{c},L_{d}\rightarrow 0$ we can offer an \textbf{a priori relation} between $\me$, $\lle$, $\mh$ and $\lh$ on the one side and the effective macroscopic elastic parameters $\lm$ and $\mm$ on the other side that we call  \textbf{macroscopic consistency condition} (see \cite{barbagallo2016transparent} for the fully anisotropic case and \cite{neff2007geometrically} for the isotropic case)
	\begin{align}
\mm  :=\frac{\mh\,\me}{\mh+\me}\,,\qquad 2\mm+3\lm  :=\frac{\left(2\mh+3\lh\right)\left(2\me+3\lle\right)}{\left(2\mh+3\lh\right)+\left(2\me+3\lle\right)}\,. \label{eq:Relation}	
\end{align} 	
For $\mh\rightarrow\infty$ we recover the \textbf{Cosserat model or micropolar model} which means that $\p \in \so(3)$ and for $L_{c}\rightarrow 0$ we obtain classical linear elasticity with $\mm,$ $\lm$ from \eqref{eq:Relation}.

For comparison, the standard isotropic Mindlin-Eringen model with $\mc>0$ and curvature energy depending on $\lVert\nablap\rVert^2$ tends to a \textbf{second gradient model} when $\me,\mc\rightarrow\infty$.

The dynamical formulation is obtained  defining the kinetic and strain energy densities of the considered mechanical system and postulating a stationary action principle. For this, we introduce a micro-inertia density contribution:
\begin{align}
J\left(u_{,t},\p_{,t}\right) & {\displaystyle \:=\frac{1}{2}\rho\left\Vert u_{,t}\right\Vert ^{2}+\frac{1}{2}\eta\left\Vert \p_{,t}\right\Vert ^{2}},
\end{align}
where $\eta$ is the scalar \textbf{micro-inertia density} and $\rho$ is the scalar \textbf{mean density}.

For us it is not at all surprising that the combination of $\Curl$ and $\Div$ in the curvature contribution at positive Cosserat couple modulus behaves similarly as does the full-micro gradient model. This is understandable since after integration and imposing boundary conditions we have the well-known inequality  \cite{neittaanmaki1984validity}:
\begin{align}
	\exists C^{+}>0\qquad \forall \p \in C^{\infty}_{0}(\Omega,\R^{3\times3}):\quad\int_{\Omega}\lVert\Curl\p\rVert^{2} +\lVert\Div\p\rVert^{2}dx \geq C^{+}(\Omega)\int_{\Omega}\lVert\nablap\rVert^{2}dx. \label{DiVCurlInequal}
\end{align}
Equation \eqref{DiVCurlInequal} means that $\lVert\Curl\p\rVert^{2}$ and $\lVert\Div\p\rVert^{2}$ considered point-wise are not equivalent to the full gradient term $\lVert\nablap\rVert^{2}$, but they become so after integration. Therefore, the $\Curl$-$\Div$-model effectively controls all first derivatives of $\p$. In consequence, the dispersion relations are similar, as can clearly be seen comparing Figures \ref{MindNon} and \ref{MindVan} with Figures \ref{DivCurlNon} and \ref{DivCurlVan}.

It should also be remarked that the well-posedness of the $\Div$-model ($L_{c}=0$) needs a strictly positive Cosserat couple modulus $\mc>0$ since an inequality of the type:
\begin{align}
	\exists C^{+}>0\qquad \forall \p \in C^{\infty}_{0}(\Omega,\R^{3\times3}):\quad\int_{\Omega}\lVert\sym\p\rVert^{2} +\lVert\Div\p\rVert^{2}dx \geq C^{+}(\Omega)\int_{\Omega}\lVert\p\rVert^{2}dx +\lVert\Div\p\rVert^{2}dx \label{SymDivInequal}
\end{align}
is \textbf{not true}. Then for $\mc>0$, there is no need for any additional inequality since the elastic energy density bounds a priori
\begin{align}
	\int_{\Omega}\lVert\p\rVert^{2} +\lVert\Div\p\rVert^{2}dx.
\end{align}
Therefore, the corresponding suitable space is a tensor-valued $H(\Div)$-Sobolev-space.

Both expressions  $\Div\p$ and $\Curl\p$ can be used to formulate a complete anisotropic curvature energy. This is possible since $\Div\p$ and $\Curl\p$ are not arbitrary collections of partial derivatives of $\p$ but satisfy the transformation laws:
\begin{align}
	\Curl_{\xi}\p^{\#}(\xi)&= Q\left[\Curl_{x}\p(x)\right]Q^{T},\qquad \xi=Q^{T}x,\qquad\mathrm{where}\ \p^{\#}(\xi):=Q\p(Q^T \xi)\,Q^{T}, \\
	\Div_{\xi}\p^{\#}(\xi)&= Q\left[\Div_{x}\p(x)\right], \nonumber
\end{align}
with respect to simultaneous rigid rotations $Q$ of the spatial and referential frame \cite[eq. (4.29) ]{munch2016some}. Therefore we may make the ansatz:
\begin{align}
	W(\nablap)=&\,W_{\Curl}(\Curl\p)+W_{\Div}(\Div\p)\\=&\,\frac{\mLc}{2}\langlenew \Lsym_{\mathrm{aniso}}\X\Curl\p,\Curl\p\ranglenew_{\R^{3\times3}}+\frac{\mLc}{2}\langlenew \Csecond_{\mathrm{aniso}}\X\Div\p,\Div\p\ranglenew_{\R^{3}},\nonumber
\end{align}
where $\Lsym_{\mathrm{aniso}}:\R^{3\times3}\rightarrow\R^{3\times3}$ is a $4^{th}$ order tensor with in general 45 independent coefficients and $\Csecond_{\mathrm{aniso}}:\R^{3}\rightarrow\R^{3}$ (for isotropy $\Csecond_{\mathrm{aniso}}$ has just 1 parameter \cite{barbagallo2016transparent}). In case of isotropy this can be significantly reduced to:
\begin{align}
	W(\nablap)=\frac{\mLc}{2}\left[\alpha_{1}\,\lVert\dev\sym\Curl\p\rVert^2+\alpha_{2}\,\lVert\skew\Curl\p\rVert^2+\frac{\alpha_{3}}{3}\left(\tr\Curl\p\right)^2+\alpha_{4}\,\lVert\Div\p\rVert^2\right].
\end{align}

\subsection{Governing equations}

The Lagrangian density $\mathscr{L}$ for the augmented relaxed model  is defined as follows:
\begin{align}
\mathscr{L}\left(u_{,t},\p_{,t},\nablau,\p,\Curl\p,\Div\p\right)=J 
\left(u_{,t},\p_{,t}\right)-W\left(\nablau,\p,\Curl\p,\Div\p\right).
\end{align}
In order to find the strong equations of motion we have to perform the first variation of the action functional
\begin{align}
\mathscr{A}\left[\left(u,\p\right)\right]:=\int_{I}\int_{\Omega}\mathscr{L}\left(u_{,t},\p_{,t},\nablau,\p,\Curl\p,\Div\p\right)dx\, dt,
\end{align}
where $I=[a,b]$ is the time interval during which we observe the motion of our system. For the kinetic part we compute
\begin{align}
\delta\int_{I}\int_{\Omega}J\left(u_{,t},\p_{,t}\right)dx\, dt & =\int_{I}\int_{\Omega}\left[D_{u_{,t}}J\left(u_{,t},\p_{,t}\right)\x\delta u_{,t}+D_{\p_{,t}}J\left(u_{,t},\p_{,t}\right)\x\delta\!\p_{,t}\right]dx\, dt\\\nonumber
\\\nonumber
& =\int_{I}\int_{\Omega}\frac{1}{2}\left[D_{u_{,t}}\left(\rho\langlenew u_{,t},u_{,t}\ranglenew \right)\x\delta u_{,t}+D_{\p_{,t}}\left(\eta\langlenew \p_{,t},\p_{,t}\ranglenew \right)\x\delta\!\p_{,t}\right]dx\, dt\\\nonumber
\\\nonumber
& =\int_{I}\int_{\Omega}\left[\rho\langlenew u_{,t},\delta u_{,t}\ranglenew +\eta\langlenew P_{,t},\delta P_{,t}\ranglenew \right]dx\, dt\\\nonumber
\\\nonumber
& =\rho\int_{\Omega}\left(\left.\langlenew u_{,t},\delta u\ranglenew \right|_{a}^{b}-\int_{I}\langlenew  u_{,tt},\delta u\ranglenew \, dt\right)dx+\eta\int_{\Omega}\left(\left.\langlenew P_{,t},\delta P\ranglenew \right|_{a}^{b}-\int_{I}\langlenew P_{,tt},\delta P\ranglenew \, dt\right)dx.
\end{align}
So considering only the bulk part we find
\begin{align}
\int_{\Omega}\int_{I}\langlenew -\,\rho\, u_{,tt},\delta u\ranglenew \, dt\, dx+\int_{\Omega}\int_{I}\langlenew -\,\eta\, P_{,tt},\delta P\ranglenew \, dt\, dx. \label{VarKin}\end{align}
For the potential part we find
\begin{align}
\delta\int_{I}\int_{\Omega} W dx\, dt =\int_{I}\int_{\Omega}\left[\langlenew D_{\,\nabla u}W,\delta\nabla u\ranglenew +\langlenew D_{\p}W,\delta\!\p\ranglenew +\langlenew D_{\Curl\p}W,\delta\Curl\p\ranglenew +\langlenew D_{\,\Div\p}W,\delta\,\Div\p\ranglenew \right]dx\, dt.
\end{align}
Having already evaluated the part $\langlenew D_{\,\nablau}W,\delta\nabla u\ranglenew +\langlenew D_{\p}W,\delta\!\p\ranglenew +\langlenew D_{\Curl\p}W,\delta\Curl\p\ranglenew $
in \cite{madeo2015wave}, we perform the explicit calculation only for the term in
$\Div\p$. So we have 
\begin{align}
\delta\int_{I}\int_{\Omega}\frac{\mLd}{2}\left\Vert \Div\p\right\Vert ^{2}dx\, dt & =\int_{I}\int_{\Omega}\frac{\mLd}{2}\,\delta\left\Vert \Div\p\right\Vert ^{2}dx\, dt =\int_{I}\int_{\Omega}\mLd\langlenew \Div\p,\delta\,\Div\p\ranglenew \, dx\, dt
 \\\nonumber &=\int_{I}\int_{\Omega}\mLd\langlenew \Div\p,\Div\,\delta\!\p\ranglenew \, dx\, dt.
\end{align}
with\footnote{Here and in the sequel $\langle\cdot,\cdot\rangle$ denotes the scalar product between two tensor of orders greater than one (e.g. $\langlenew A,B \ranglenew=A_{ij}B_{ij}$). Moreover a central dot stands for the simple contraction between two tensors of an order greater than one. For example $(A\x v)_{i}=A_{ij}v_{j}$. Finally we use Einstein convention of sum over repeated indexes if not differently specified.}
\begin{align}
\langlenew \Div\p,\Div\,\delta\!\p\ranglenew =\Div\left(\Div\p\x\delta\!\p\right)-\langlenew \nabla\,\Div\p,\delta\!\p\ranglenew 
\end{align}
that in index notation is
\begin{align}
\p_{ij,j}\delta\!\p_{ih,h}=\left(\p_{ij,j}\delta\!\p_{ih}\right)_{,h}-\p_{ij,jh}\delta\!\p_{ih}\,,
\end{align}
we integrate by parts and find that
\begin{align}
\delta\int_{I}\int_{\Omega}\frac{\mLd}{2}\left\Vert \Div\p\right\Vert ^{2}dx\, dt & =\int_{I}\int_{\Omega}\mLd\left[\,\Div\left(\,\Div\p\cdot\delta\!\p\,\right)-\langlenew \nabla\,\Div\p,\delta\!\p\ranglenew \right] dx\, dt \label{VarDiv}\\\nonumber
& =\int_{I}\int_{\partial\Omega}\mLd\langlenew \Div\p\x\delta\!\p,\,n\ranglenew \, ds\,dt +\int_{I}\int_{\Omega}\langlenew -\,\mLd\,\nabla\,\Div\p,\delta\!\p\ranglenew \, dx\, dt,
\end{align}
where $n$ is the unit normal field to the boundary. 
Considering only the kinetic energy associated to $P$ and the potential energy
related to $\Div\p$ we have
\begin{align}
\int_{I}\int_{\Omega}\left(\frac{1}{2}\eta\left\Vert \p_{,t}\right\Vert ^{2}-\frac{\mLd}{2}\left\Vert \Div\p\right\Vert ^{2}\right)dx\, dt
\end{align}
and, with reference to equations \eqref{VarKin} and \eqref{VarDiv}, the bulk part of the first variation is 
\begin{align}
\int_{I}\int_{\Omega}\left(\langlenew -\,\eta\, P_{,tt},\delta P\ranglenew -\langlenew -\,\mLd\,\nabla\,\Div\p,\delta\!\p\ranglenew \right)dx\, dt=\int_{I}\int_{\Omega}\langlenew -\,\eta\, P_{,tt}+\mLd\,\nabla\,\Div\p,\delta\!\p\ranglenew \, dx\, dt.
\end{align}
Altogether, see also \cite{madeo2015wave}, the strong equations in the bulk are
\begin{align}
\rho\, u_{,tt} & =\Div\left[2\,\me\,\sym\left(\nablau-\p\right)+\lle\,\tr\left(\nablau-\p\right)\id+2\,\mc\,\skew\left(\nablau-\p\right)\right],\nonumber
\\\nonumber
\\\label{eq:strong equations}
\eta\,\p_{,tt} & =2\,\me\,\sym\left(\nablau-\p\right)+\lle\,\tr\left(\nablau-\p\right)\id+2\,\mc\,\skew\left(\nablau-\p\right)\\\nonumber
\\\nonumber
& \quad-2\,\mh\,\sym\p-\lh\,\tr\left(\!\p\right)\id-\mu\, L_{c}^{2}\,\Curl\,\Curl\p+\hspace{-0.3cm}\underbrace{\mLd\,\nabla\,\Div\p}_{\text{new augmented term}}\hspace{-0.3cm}.
\end{align}

In our study of wave propagation in micromorphic media we limit ourselves to the case of \textbf{plane waves} traveling in an \textbf{infinite domain}. We suppose that the space dependence of all introduced kinematic fields are limited to the component $x_{1}$ of $x$ which is also the direction of propagation of the wave. Therefore we look for solutions of \eqref{eq:strong equations} in the form:
\begin{align}
u(x,t)=\alpha \, e^{i \left(k\, x_{1}-\,\omega \,t\right)}\,,\ \alpha\in\R^{3}\,,\qquad\qquad \p(x,t) =\beta\, e^{i \left(k \,x_{1}-\,\omega \,t\right)}\,,\ \beta\in\R^{3\times3}\,.\label{eq:PlaneWave}
\end{align}

\subsection{Decomposition of the equations of motion}

Considering the system of PDEs found in \eqref{eq:strong
equations}, we can rewrite this system in a fashion more convenient
for the study of the propagation of plane waves in a homogeneous
isotropic medium. Our approach consists always in projecting the found
relations in the three orthogonal sub vector spaces $\Symtre\cap\sll,\soo,\langlenew \id\ranglenew $. In this way, a tensor $X\in\R^{3\times3}$ is uniquely written by means of the Cartan-Lie decomposition as:
\begin{align}
X=\devsym\left(X\right)+\skew\left(X\right)+\frac{1}{3}\tr\left( X\right)\id 
\end{align} 
where
\begin{gather}
\devsym\left(X\right)  =\begin{pmatrix}X^{D} & X_{\left(12\right)} & X_{\left(13\right)}\\
\\
X_{\left(12\right)} & X_{2}^{D} & X_{\left(23\right)}\\
\\
X_{\left(13\right)} & X_{\left(23\right)} & X_{3}^{D}
\end{pmatrix},\quad
\skew\left(X\right) =\begin{pmatrix}0 & X_{\left[12\right]} & X_{\left[13\right]}\\
\\
-X_{\left[12\right]} & 0 & X_{\left[23\right]}\\
\\
-X_{\left[13\right]} & -X_{\left[23\right]} & 0
\end{pmatrix},\quad\nonumber
\\
\\
\frac{1}{3}\tr\left( X\right)\id = X^{S} \id \nonumber
,
\end{gather}
in which we set
\begin{align}
 X^{S}&=\frac{1}{3}\left(X_{11}+X_{22}+X_{33}\right), \  &X_{\left[12\right]}&= \frac{1}{2}\left(X_{12}-X_{21}\right), & X_{\left(12\right)}&={\displaystyle \frac{1}{2}\left(X_{12}+X_{21}\right),}\nonumber\\\nonumber
\\
X^{D}&=X_{11}-X^{S}, & X_{\left[13\right]}&= \frac{1}{2}\left(X_{13}-X_{31}\right), \  &X_{\left(13\right)}&=\frac{1}{2}\left(X_{13}+X_{31}\right),\\\nonumber
\\\nonumber
X_{\alpha}^{D}&=X_{\alpha\alpha}-X^{S}, &X_{\left[23\right]}&={\displaystyle \frac{1}{2}\left(X_{23}-X_{32}\right),} \  &X_{\left(23\right)} &={\displaystyle \frac{1}{2}\left(X_{23}+X_{32}\right).}
\end{align}
The components $X_{2}^{D}$ and $X_{3}^{D}$ are not independent,
but are related by the following relation
\begin{align}
X_{2}^{D}-X_{3}^{D}=X^{V}=P_{22}-P_{33}.
\end{align}
In this way, applying the Cartan-Lie decomposition to the tensor $X=\sym \p$ in the first equation and to all the tensors appearing in the second one, the equations \eqref{eq:strong equations} can be written as follows
\begin{align}
\rho\, u_{,tt} & =\Div\left[2\,\me\,\sym\left(\nablau-\p\right)+\lle\,\tr\left(\nablau-\p\right)\id+2\,\mc\,\skew\left(\nablau-\p\right)\right],\nonumber
\\\nonumber\\
\eta\,\left(\devsym\p_{,tt}\right) & =2\,\me\,\devsym\left(\nablau-P\right)-2\,\mh\,\devsym\p-\mu\, L_{c}^{2}\,\devsym\left(\Curl\,\Curl\p\right)\nonumber\\\nonumber&\qquad+\mLd\,\dev\sym\left(\nabla\,\Div\p\right),
\\ \\\nonumber
\eta\,\left(\skew\p_{,tt}\right)&  =2\,\mc\,\skew\left(\nablau-\p\right)-\mu\, L_{c}^{2}\,\skew\left(\Curl\,\Curl\p\right)+\mu\, L_{d}^{2}\,\skew\left(\nabla\,\Div\p\right),\\\nonumber
\\\nonumber
\eta\,\frac{1}{3}\tr\left(\!\p_{,tt}\right)\id & =\left(\frac{2\,\me+3\,\lle}{3}\right)\tr\left(\nablau-P\right)\id-\left(\frac{2\,\mh+3\,\lh}{3}\right)\tr\left(P\right)\id\\\nonumber
&\qquad-\mu\, L_{c}^{2}\,\frac{1}{3}\tr\left(\Curl\,\Curl\p\right)\id+\mu\, L_{d}^{2}\,\frac{1}{3}\tr\left(\nabla\,\Div\p\right)\id,
\end{align}
where we have only five independent equations for the $\devsym$-part,
three independent equations for the $\skew$-part and one independent
equation for the spherical part. 

If we demand that the kinematic fields $u$ and $P$ are plane waves in the $x_{1}$ direction as indicated in \eqref{eq:PlaneWave}, we have equivalently the following expressions in index notation:
\begin{align}
u_{i}\left(x,t\right)&=u_{i}\left(x_{1},t\right)=\alpha_{i}\, e^{i\left(kx_{1}-\,\omega t\right)},\vspace{0.2cm}
\\\nonumber
P_{ij}\left(x,t\right)&=P_{ij}\left(x_{1},t\right)=\beta_{ij}\, e^{i\left(kx_{1}-\,\omega t\right)}.
\end{align}
In this way, it is easy to derive the expression in components
of the projected equations. With respect to the article \cite{madeo2015wave},
we have to explicitly calculate only the new part in $\nabla\,\Div\p$.
We have that 
\[
\nabla\,\Div\p=\nabla\,\Div\,\devsym\,\p+\nabla\,\Div\,\skew\,\p+\nabla\,\Div\left(\frac{1}{3}\tr\left(\!\p\right)\id\right),
\]
so 
\begin{align}
\devsym\,\nabla\,\Div\p & =\devsym\left(\nabla\,\Div\,\devsym\,\p+\nabla\,\Div\,\skew\,\p+\nabla\,\Div\,\frac{1}{3}\tr\left(\!\p\right)\id\right),\nonumber\vspace{0.2cm}
\\
\skew\,\nabla\,\Div\p & =\skew\left(\nabla\,\Div\,\devsym\,\p+\nabla\,\Div\,\skew\,\p+\nabla\,\Div\,\frac{1}{3}\tr\left(\!\p\right)\id\right),\vspace{0.2cm}
\\\nonumber
\frac{1}{3}\tr\left(\,\nabla\,\Div\p\right)\id & =\frac{1}{3}\tr\left(\nabla\,\Div\,\devsym\,\p+\nabla\,\Div\,\skew\,\p+\nabla\,\Div\,\frac{1}{3}\tr\left(\!\p\right)\id\right)\id,
\end{align}
and finally, using the fact that $\p$ is assumed to depend only on the scalar space variable $x_{1}$, we obtain 
\begin{align}
\devsym\,\nabla\,\Div\p & =\begin{pmatrix}\frac{2}{3}P_{,11}^{D}+\frac{2}{3}P_{,11}^{S} & \frac{1}{2}P_{\left(12\right),11}-\frac{1}{2}P_{\left[12\right],11} & \frac{1}{2}P_{\left(13\right),11}-\frac{1}{2}P_{\left[13\right],11}\vspace{0.2cm}
\\
\frac{1}{2}P_{\left(12\right),11}-\frac{1}{2}P_{\left[12\right],11} & -\frac{1}{3}P_{,11}^{D}-\frac{1}{3}P_{,11}^{S} & 0\vspace{0.2cm}
\\
\frac{1}{2}P_{\left(13\right),11}-\frac{1}{2}P_{\left[13\right],11} & 0 & -\frac{1}{3}P_{,11}^{D}-\frac{1}{3}P_{,11}^{S}
\end{pmatrix},\nonumber\\\nonumber
\\
\skew\,\nabla\,\Div\p & =\frac{1}{2}\begin{pmatrix}0 & -P_{\left(12\right),11}+P_{\left[12\right],11} & -P_{\left(13\right),11}+P_{\left[13\right],11}\vspace{0.2cm}\\
P_{\left(12\right),11}-P_{\left[12\right],11} & 0 & 0\vspace{0.2cm}\\
P_{\left(13\right),11}-P_{\left[13\right],11} & 0 & 0
\end{pmatrix},\\\nonumber
\\\nonumber
\frac{1}{3}\tr\left(\,\nabla\,\Div\p\right)\id & 
=\frac{1}{3}\left(P_{,11}^{D}+P_{,11}^{S}\right)\id
.
\end{align}
Introducing the quantities\footnote{Due to the chosen values of the parameters, which are supposed to	satisfy (\ref{DefPos}), all the introduced characteristic velocities and frequencies are real. Indeed it can be checked that the condition $\left(3\,\lambda_{e}+2\,\mu_{e}\right)>0$ together with 	the condition $\mu_{e}>0$ imply $\left(\lambda_{e}+2\mu_{e}\right)>0$.}
\begin{align}
c_{m}&=\sqrt{\frac{\mLc}{\eta}},\quad &c_{d}&=\sqrt{\frac{\mLd}{\eta}},\quad &c_{s}&=\sqrt{\frac{\mu_{e}+\mu_{c}}{\rho}},\nonumber\\\nonumber\\ \label{Definitions}
 c_{p}&=\sqrt{\frac{\lambda_{e}+2\mu_{e}}{\rho}},\quad
&\omega_{s}&=\sqrt{\frac{2\left(\me+\mh\right)}{\eta}},\quad
&\omega_{p}&=\sqrt{\frac{2\left(\me+\mh\right)+3\left(\lle+\lh\right)}{\eta}},
\\\nonumber\\\nonumber
\omega_{r}&=\sqrt{\frac{2\mu_{c}}{\eta}},\quad
&\omega_{l}&=\sqrt{\frac{\lh+2\mh}{\eta}},\quad
&\omega_{t}&=\sqrt{\frac{\mh}{\eta}},
\end{align}
the equations can be written as:
\begin{itemize}
	\item a set of three equations only involving longitudinal quantities:
\begin{align*}
\ddot{u}_{1} & =c_{p}^{2}u_{1,11}-\frac{2\mu_{e}}{\rho}\, P_{,1}^{D}-\frac{3\lambda_{e}+2\mu_{e}}{\rho}\, P_{,1}^{S}\,,\vspace{0.4cm}
\\
\ddot{P}^{D} & =\frac{4}{3}\,\frac{\mu_{e}}{\eta}\,u_{1,1}+\frac{1}{3}\,c_{m}^{2}\,P_{,11}^{D}-\frac{2}{3}\,c_{m}^{2}P_{,11}^{S}-\omega_{s}^{2}\,P^{D}+\underbrace{\frac{2}{3}\,c_{d}^{2}\,P_{,11}^{D}+\frac{2}{3}\,c_{d}^{2}\,P_{,11}^{S}}_{\text{new augmented terms}}\,,\vspace{0.4cm}
\\
\ddot{P}^{S} & =\frac{3\lambda_{e}+2\mu_{e}}{3\eta}\,u_{1,1}-\frac{1}{3}\,c_{m}^{2}P_{,11}^{D}+\frac{2}{3}\,c_{m}^{2}P_{,11}^{S}-\omega_{p}^{2}\,P^{S}+\underbrace{\frac{1}{3}\,c_{d}^{2}\,P_{,11}^{D}+\frac{1}{3}\,c_{d}^{2}\,P_{,11}^{S}}_{\text{new augmented terms}}\,,
\end{align*}
	\item two sets of three equations only involving transverse quantities in
	the $\xi$-th direction, with $\xi=2,3$:
\begin{align*}
\ddot{u}_{\xi} & =c_{s}^{2}u_{\xi,11}-\frac{2\mu_{e}}{\rho}\, P_{\left(1\xi\right),1}+\frac{\eta}{\rho}\,\omega_{r}^{2}P_{\left[1\xi\right],1},\vspace{0.4cm}
\\
\ddot{P}_{\left(1\xi\right)} & =\frac{\mu_{e}}{\eta}\,u_{\xi,1}+\frac{1}{2}\,c_{m}^{2}\,P_{(1\xi)}{}_{,11}+\frac{1}{2}\,c_{m}^{2}\,P_{\left[1\xi\right],11}-\omega_{s}^{2}\,P_{(1\xi)}+\underbrace{\frac{1}{2}\,c_{d}^{2}\,P_{\left(1\xi\right),11}-\frac{1}{2}\,c_{d}^{2}\,P_{\left[1\xi\right],11}}_{\text{new augmented terms}},\vspace{0.4cm}
\\
\ddot{P}_{\left[1\xi\right]} & =-\frac{1}{2}\,\omega_{r}^{2}\,u_{\xi,1}+\frac{1}{2}\,c_{m}^{2}\,P_{(1\xi),11}+\frac{1}{2}\,c_{m}^{2}P_{\left[1\xi\right]}{}_{,11}-\omega_{r}^{2}\,P_{\left[1\xi\right]}-\underbrace{\frac{1}{2}\,c_{d}^{2}\,P_{\left(1\xi\right),11}+\frac{1}{2}\,c_{d}^{2}\,P_{\left[1\xi\right],11}}_{\text{new augmented terms}},
\end{align*}
\end{itemize}
\begin{itemize}
	\item One equation only involving the variable $P_{\left(23\right)}$:\begin{align*}
	\ddot{P}_{\left(23\right)}=-\omega_{s}^{2}P_{\left(23\right)}+c_{m}^{2}P_{\left(23\right),11},\label{Shear}
\end{align*}
	\item One equation only involving the variable $P_{\left[23\right]}$ :
	\[
	\ddot{P}_{\left[23\right]}=-\omega_{r}^{2}P_{\left[23\right]}+c_{m}^{2}P_{\left[23\right],11},\label{Rotations23}
	\]
	
	\item One equation only involving the variable $P^{V}$:
	\[
	\ddot{P}^{V}=-\omega_{s}^{2}P^{V}+c_{m}^{2}P_{,11}^{V}.\label{VolumeVariation}
	\]
\end{itemize}
 
\section{Particularization for specific energies}
In what follows we will present the results obtained with particular energies and the numerical values of the elastic coefficients are chosen as in Table \ref{ParametersValues} if not differently specified.
		\begin{table}[H]
	\begin{centering}
		\begin{tabular}{|c|c|c|}
			\hline 
			Parameter & Value & Unit\tabularnewline
			\hline 
			\hline 
			$\me$ & $200$ & $MPa$\tabularnewline
			\hline 
			$\lle=2\me$ & 400 & $MPa$\tabularnewline
			\hline 
			$\mc=5\me$ & $ $1000 & $MPa$\tabularnewline
			\hline 
			$\mh$ & 100 & $MPa$\tabularnewline
			\hline 
			$\lh$ & $100$ & $MPa$\tabularnewline
			\hline 
			$L_{c}\ $  & $1$ & $mm$\tabularnewline
			\hline 
			$\rho$ & $2000$ & $Kg/m^{3}$\tabularnewline
			\hline 
			$\eta$ & $ $$10^{-2}$ & $Kg/m$\tabularnewline
			\hline 
		\end{tabular}\quad{}\quad{}\quad{}\quad{}%
		\begin{tabular}{|c|c|c|}
			\hline 
			Parameter & Value & Unit\tabularnewline
			\hline 
			\hline 
			$\lm$ & $82.5$ & $MPa$\tabularnewline
			\hline 
			$\mm$ & $66.7$ & $MPa$\tabularnewline
			\hline 
			$E_{\mathrm{macro}}$ & $170$ & $MPa$\tabularnewline
			\hline 
			$\nu_{\mathrm{macro}}$ & $0.28$ & $-$\tabularnewline
			\hline 
		\end{tabular}
		\par\end{centering}
	
	\caption{\label{ParametersValues}Values of the parameters 
		used in the numerical simulations (left) and corresponding values
		of the Lamé parameters and of the Young modulus and Poisson ratio as obtained with formula \eqref{eq:Relation}	
		(right).}

\end{table}

\subsection{The micromorphic model with $\rVert\Div\p\lVert^{2}$ and  $\rVert\Curl\p\lVert^{2}$ ($L_{c}=L_{d}\neq0$)\label{DIVCURL}}

We consider now the model obtained considering $L_{c}=L_{d}$ with energy:
\begin{align}
W=&\underbrace{\me\,\lVert \sym\left(\nablau-\p\right)\rVert ^{2}+\frac{\lle}{2}\left(\mathrm{tr} \left(\nablau-\p\right)\right)^{2}}_{\mathrm{{\textstyle isotropic\ elastic-energy}}}	+\hspace{-0.1cm}\underbrace{\mc\,\lVert \skew\left(\nablau-\p\right)\rVert ^{2}}_{\mathrm{\textstyle rotational\   elastic\ coupling}		}\hspace{-0.1cm} \label{eq:Ener-DivCurl}\\
& \quad 
+\underbrace{\mh\,\lVert \sym \p\rVert ^{2}+\frac{\lh}{2}\,\left(\mathrm{tr} \p\right)^{2}}_{\mathrm{{\textstyle micro-self-energy}}}
+\underbrace{\frac{\mLc}{2} \,\left(\lVert \Div \p\rVert^2+\lVert \Curl \p\rVert^2\right)}_{\mathrm{\textstyle augmented\ isotropic\ curvature}}\,.
\nonumber  
\end{align}

The dynamical equilibrium equations are:
\begin{align}
\rho\,  u_{,tt}=&\,\Div\,\sigma=\,\Div\left[2\,\me\, \sym\left(\nablau-\p\right)+2\,\mc \,\skew\left(\nablau-\p\right)+\lle\tr\left(\nablau-\p\right)\mathds{1}\right] , \nonumber \\
\eta  \p_{,tt}=&\,2\, \me \,\sym\left(\nablau-\p\right) +2 \,\mc\, \skew\left(\nablau-\p\right)+\lle\tr\left(\nablau-\p\right)\mathds{1}\label{eq:DynCurlDiv}\\&\ -\left[2\mh \X\sym \p +\lh \tr(\p)\mathds{1}\right]+\mLc \, \underbrace{\left(\nabla \left( \Div \p\right)-\Curl  \Curl \p\right)}_{ \Div\nabla  \p=\Delta\p}.\nonumber 
\end{align}
Note that the structure of the equation is equivalent to the one obtained in the standard micromorphic model with curvature $\frac{1}{2}\rVert\nablap\lVert^{2}$, see equation \eqref{eq:DynMind} in section \ref{sec:Mind}. 

\begin{figure}[H]
	\begin{centering}
		\begin{tabular}{ccccc}
			(a)\includegraphics[width=4cm]{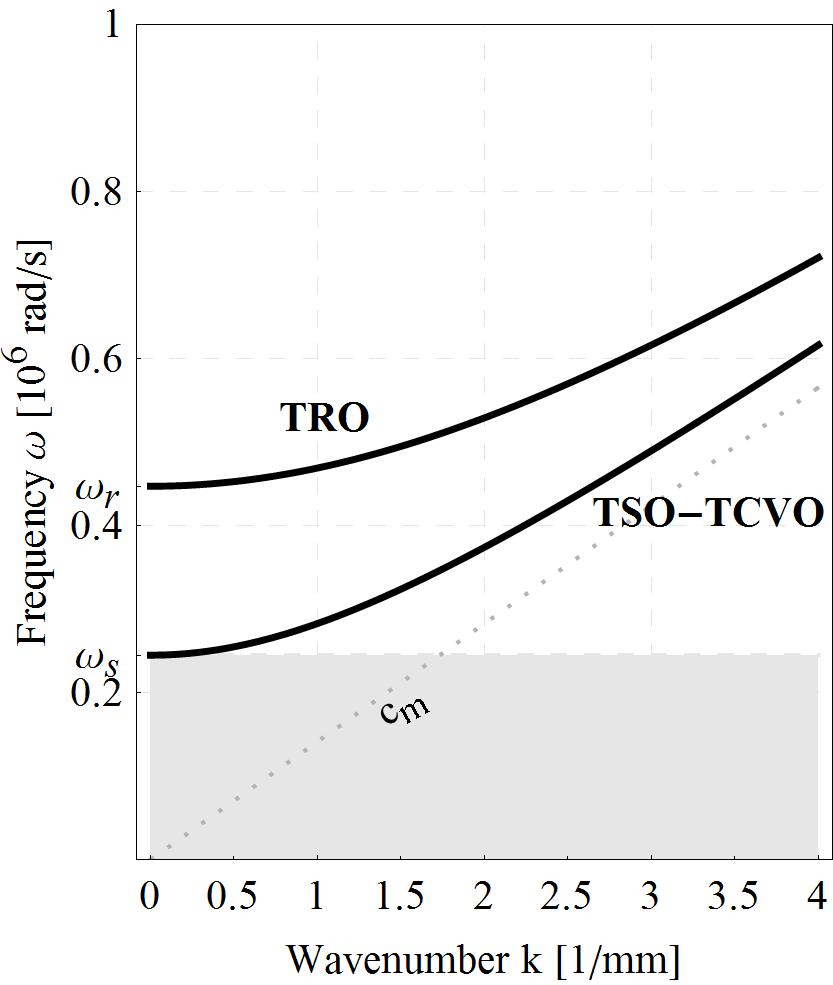}
		&\qquad(b) \includegraphics[width=4cm]{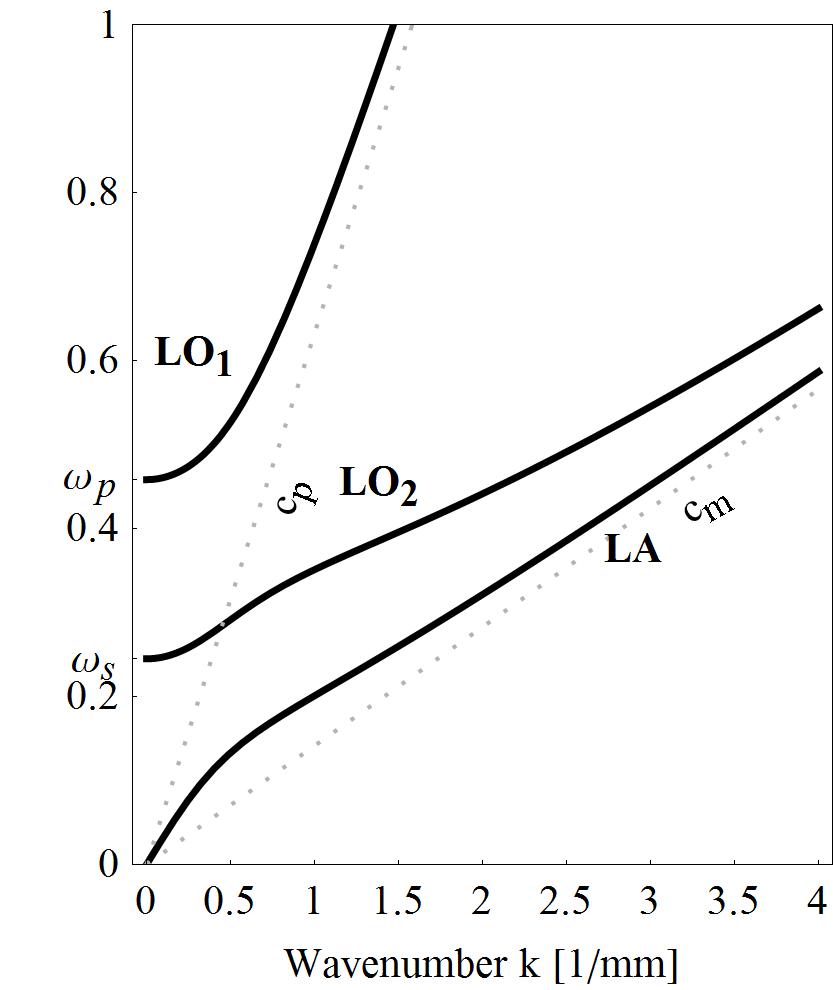} &\qquad(c) \includegraphics[width=4cm]{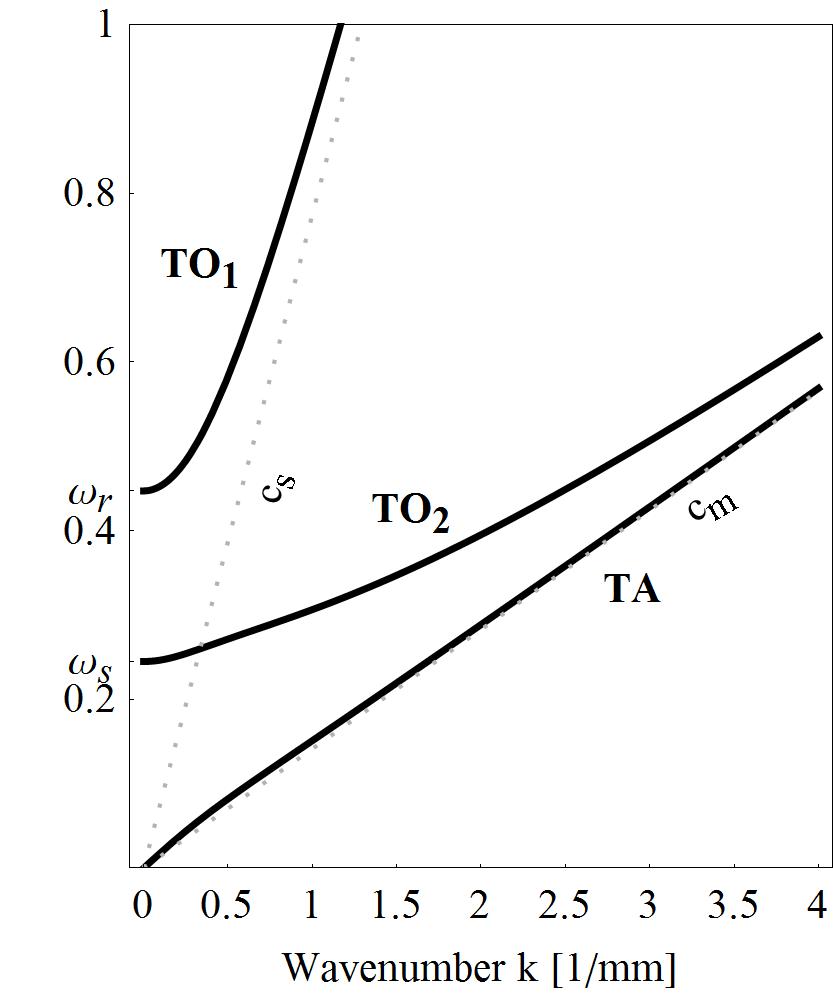} &
		\end{tabular}
		\par\end{centering}
	
	\caption{\label{DivCurlNon}Dispersion relations $\omega=\omega(k)$ for the
		\textbf{micromorphic model with $\rVert\Div\p\lVert^{2}+\rVert\Curl\p\lVert^{2}$} and non-vanishing Cosserat couple modulus $\mc>0$:  only a 	\textbf{partial band gap} on the uncoupled waves can be modeled.
	}	
\end{figure}

\begin{figure}[H]
	\begin{centering}
		\begin{tabular}{ccccc}
			(a)\includegraphics[width=4cm]{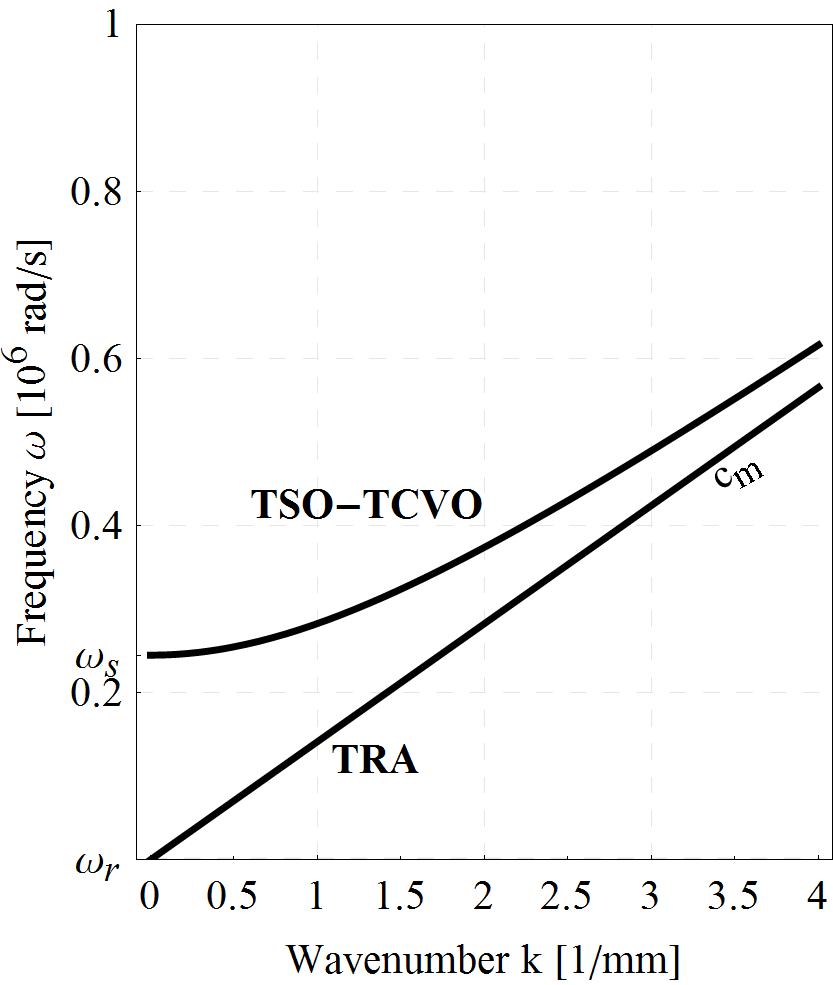}  &\qquad(b) \includegraphics[width=4cm]{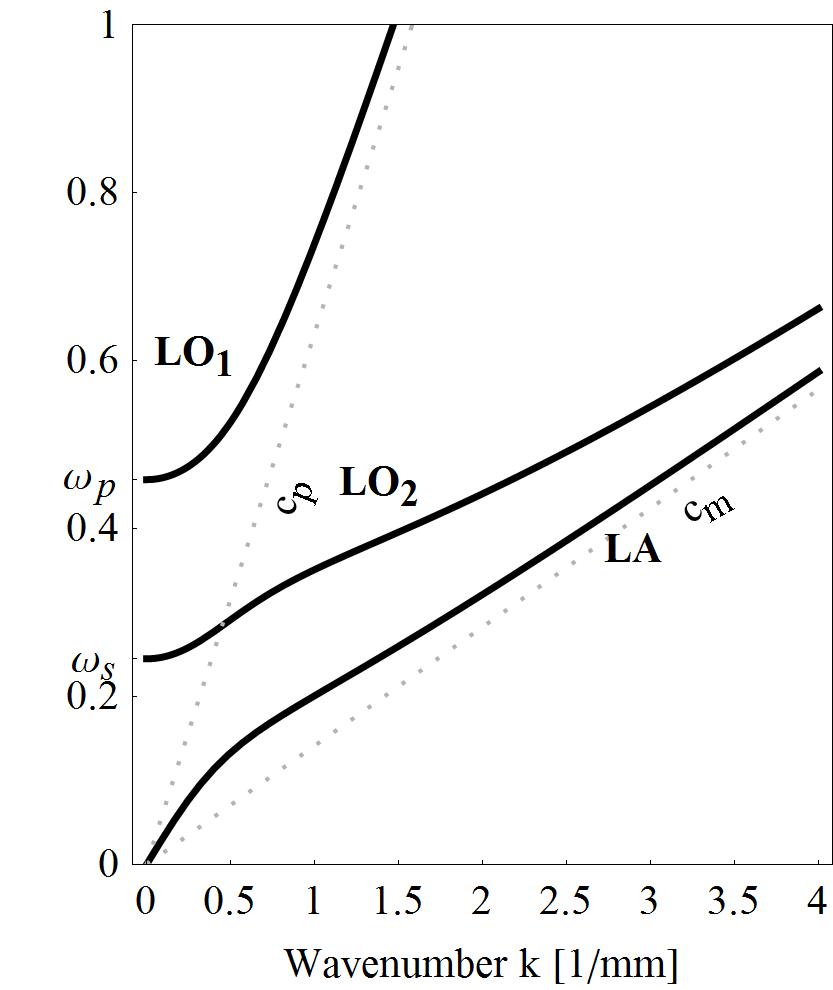} &\qquad(c) \includegraphics[width=4cm]{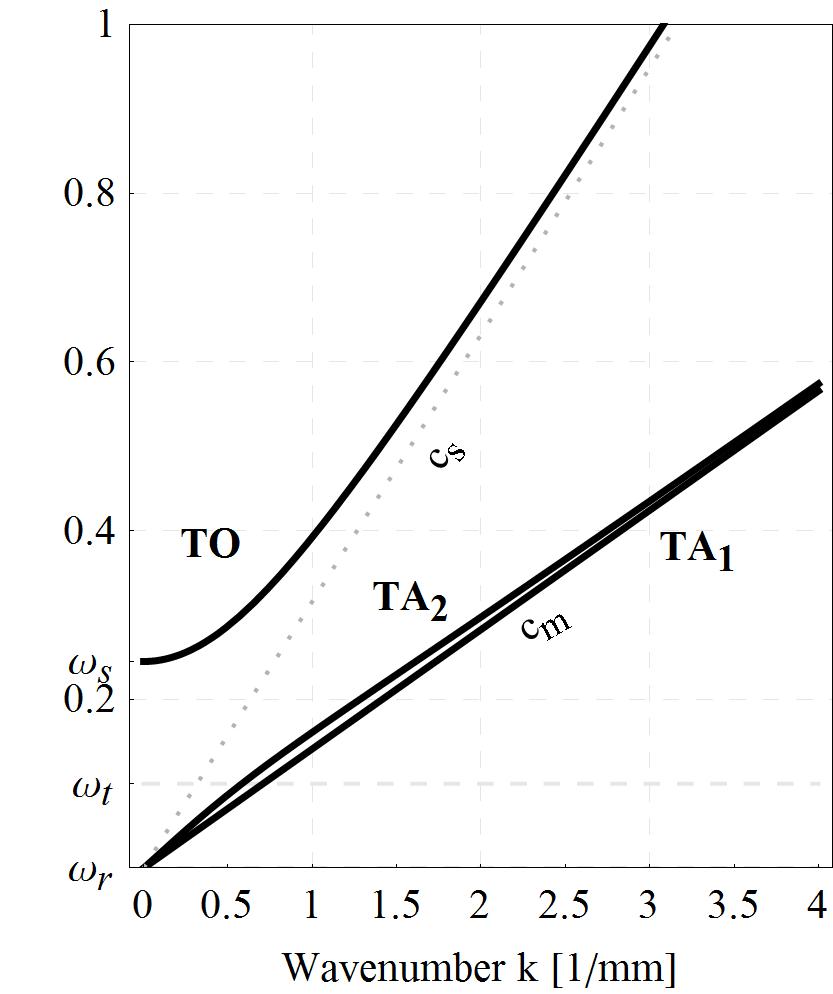} &
		\end{tabular}
		\par\end{centering}
	
	\caption{\label{DivCurlVan}Dispersion relations $\omega=\omega(k)$ for the
		\textbf{micromorphic model with $\rVert\Div\p\lVert^{2}+\rVert\Curl\p\lVert^{2}$} and vanishing Cosserat couple modulus $\mc=0$: \textbf{no band gap} at all.
	}	
\end{figure}

We present the \textbf{dispersion relations} obtained with a non-vanishing Cosserat couple modulus $\mc>0$ (Figure \ref{DivCurlNon}) and for a vanishing Cosserat couple modulus $\mc=0$ (Figure \ref{DivCurlVan}). In all the figures we consider  uncoupled waves (a), longitudinal waves (b) and
transverse waves (c). The nomenclature adopted is the following: TRO: transverse rotational optic, TSO: transverse shear optic, TCVO:
transverse constant-volume optic, LA: longitudinal acoustic, LO$_{1}$-LO$_{2}$:
$1^{st}$ and $2^{nd}$ longitudinal optic, TA: transverse acoustic, TO$_{1}$-TO$_{2}$: $1^{st}$ and $2^{nd}$ transverse optic.

We conclude that when considering the model with micromorphic medium with $\rVert\Div\p\lVert^{2}$ + $\rVert\Curl\p\lVert^{2}$ and vanishing Cosserat couple modulus $\mu_{c}$, there always exist waves which propagate inside the considered medium independently of the value of frequency even if considering separately  longitudinal, transverse and uncoupled waves. The only effect obtainable switching on the Cosserat couple modulus $\mu_{c}$ is to obtain a partial band gap for the uncoupled waves.

\subsection{The micromorphic model with only $\rVert\Div\p\lVert^{2}$ obtained  as a special case of the augmented relaxed model with $L_{c}=0$}

The isotropic micromorphic model with $\rVert\Div\p\lVert^{2}$ is obtained from the model with $\rVert\Curl\p\lVert^{2}$ and $\rVert\Div\p\lVert^{2}$ by considering $L_{c}=0$ obtaining as standard energy:
\begin{align}
W=&\underbrace{\me\,\lVert \sym\left(\nablau-\p\right)\rVert ^{2}+\frac{\lle}{2}\left(\mathrm{tr} \left(\nablau-\p\right)\right)^{2}}_{\mathrm{{\textstyle isotropic\ elastic-energy}}}	+\hspace{-0.1cm}\underbrace{\mc\,\lVert \skew\left(\nablau-\p\right)\rVert ^{2}}_{\mathrm{\textstyle rotational\   elastic\ coupling}		}\hspace{-0.1cm} \label{eq:Ener-Div}\\
& \quad 
+\underbrace{\mh\,\lVert \sym \p\rVert ^{2}+\frac{\lh}{2}\,\left(\mathrm{tr} \p\right)^{2}}_{\mathrm{{\textstyle micro-self-energy}}}
+\hspace{-0.2cm}\underbrace{\frac{\mLd}{2} \,\lVert \Div \p\rVert^2}_{\mathrm{\textstyle isotropic\ curvature}}\,.
\nonumber  
\end{align}
The dynamical equilibrium equations are:
\begin{align}
\rho\,  u_{,tt}=&\,\Div\,\sigma=\,\Div\left[2\,\me\, \sym\left(\nablau-\p\right)+2\,\mc \,\skew\left(\nablau-\p\right)+\lle\tr\left(\nablau-\p\right)\mathds{1}\right] , \nonumber \\
\eta  \p_{,tt}=&\,2\, \me \,\sym\left(\nablau-\p\right) +2 \,\mc\, \skew\left(\nablau-\p\right)+\lle\tr\left(\nablau-\p\right)\mathds{1}\label{eq:DynDiv}\\&\ -\left[2\mh \X\sym \p +\lh \tr(\p)\mathds{1}\right]+\mLd \, \nabla \left( \Div \p\right).\nonumber 
\end{align}

\begin{figure}[H]
	\begin{centering}
		\begin{tabular}{ccccc}
			(a)\includegraphics[width=4cm]{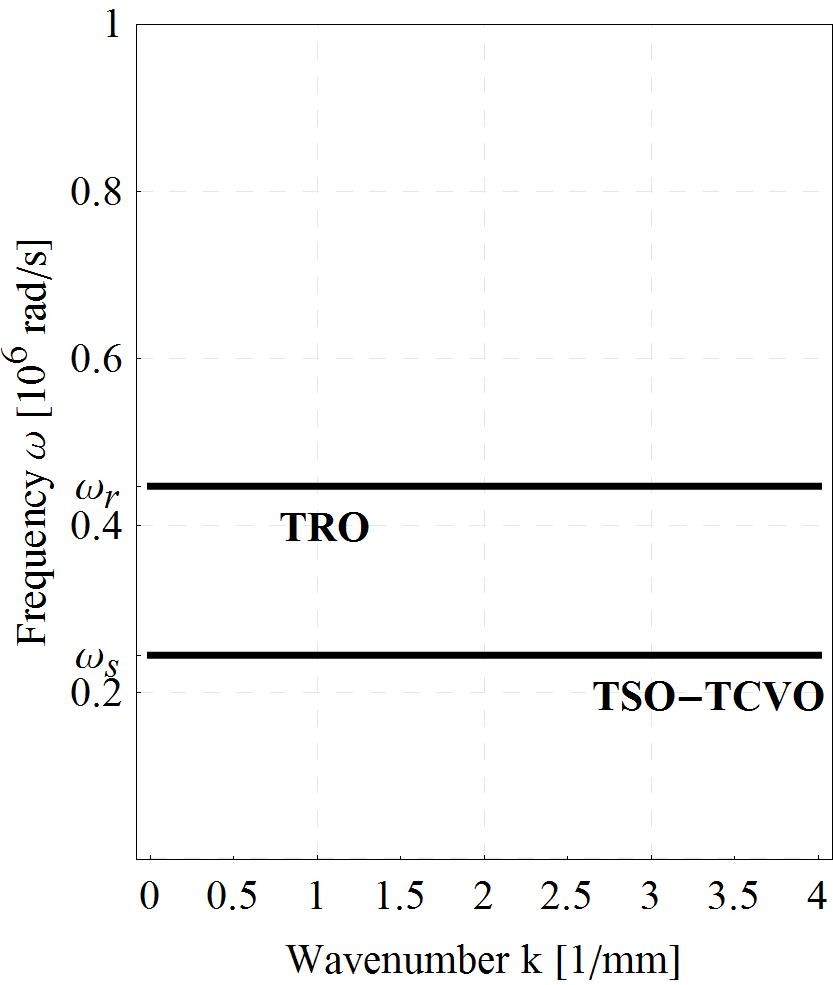}  &\qquad(b) \includegraphics[width=4cm]{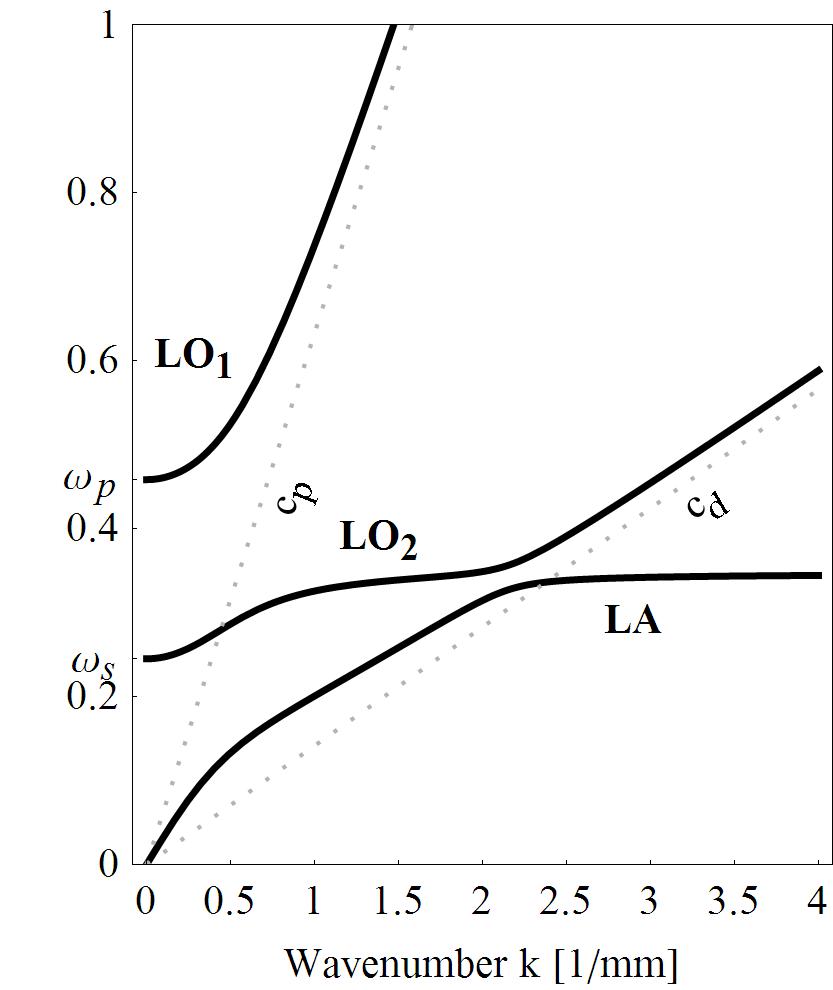} &\qquad(c) \includegraphics[width=4cm]{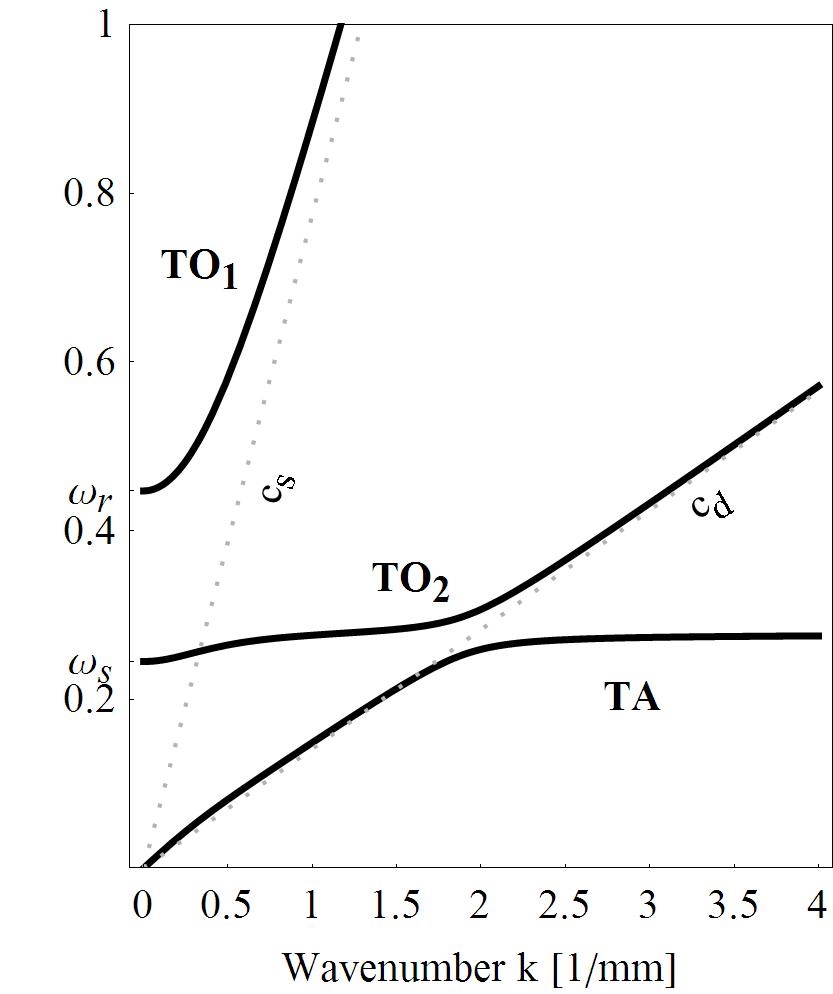} &
		\end{tabular}
		\par\end{centering}
	
	\caption{\label{DivNon}Dispersion relations $\omega=\omega(k)$ for the
		\textbf{micromorphic model  with $\rVert\Div\p\lVert^{2}$} and non-vanishing Cosserat couple modulus $\mc>0$:  no	\textbf{band gap} on the longitudinal and transverse waves can be modeled and the uncoupled waves have fixed frequencies.
	}	
\end{figure}

\begin{figure}[H]
	\begin{centering}
		\begin{tabular}{ccccc}
			(a)\includegraphics[width=4cm]{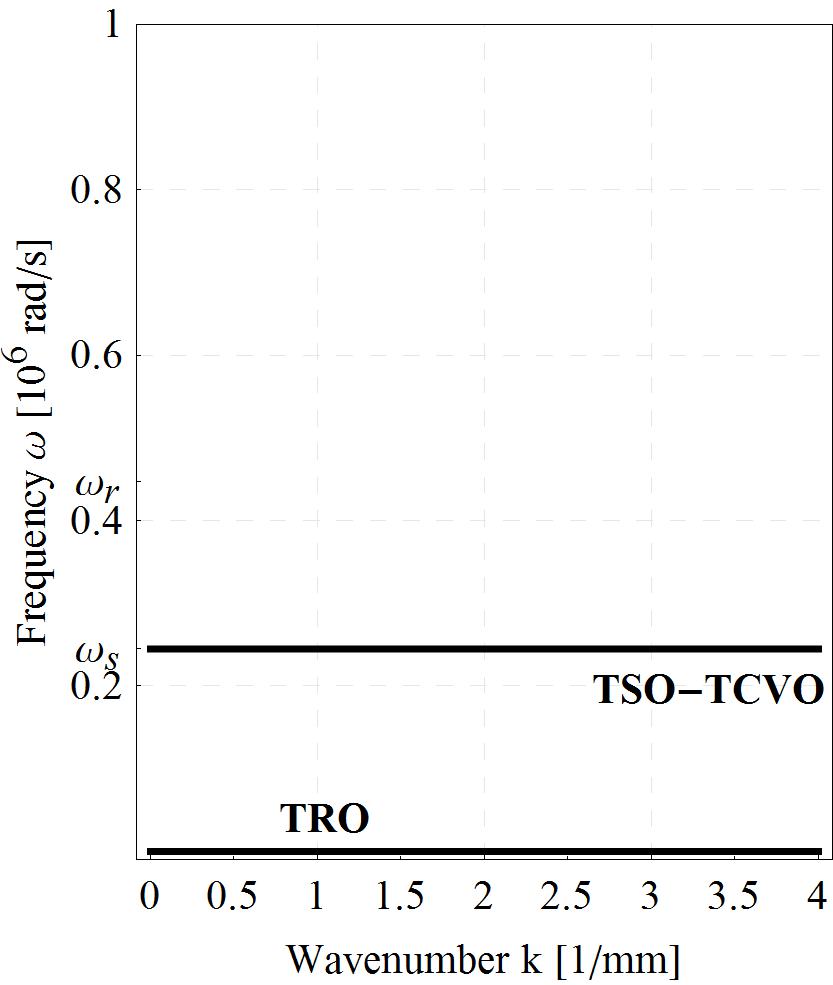}  &\qquad(b) \includegraphics[width=4cm]{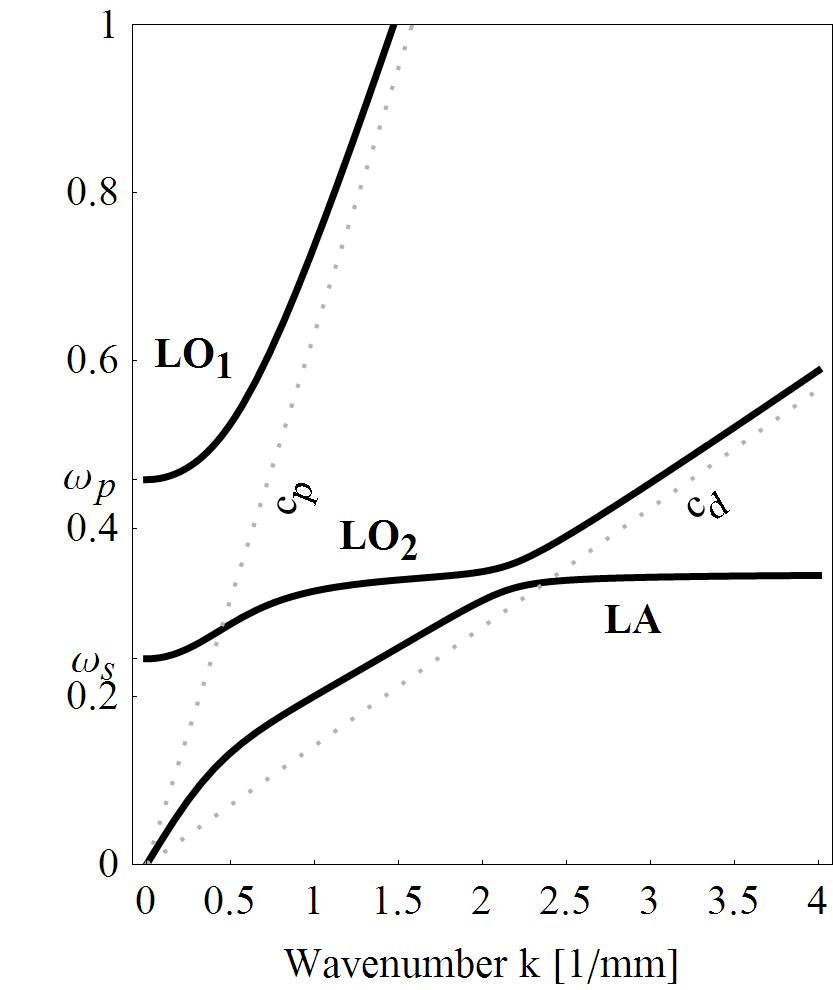} &\qquad(c) \includegraphics[width=4cm]{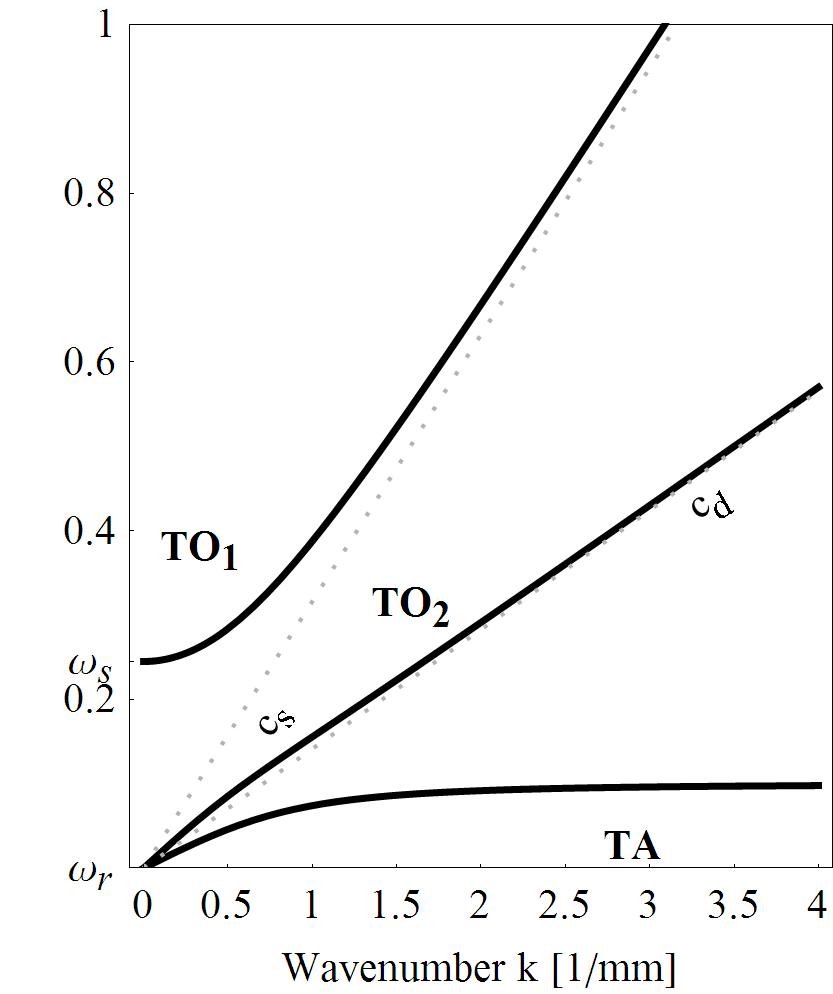} &
		\end{tabular}
		\par\end{centering}
	
	\caption{\label{DivVan}Dispersion relations $\omega=\omega(k)$ for the
		\textbf{micromorphic model  with $\rVert\Div\p\lVert^{2}$} and vanishing Cosserat couple modulus $\mc=0$:  no	\textbf{band gap} on the longitudinal and transverse waves can be modeled and the uncoupled waves have fixed frequencies.
	}	
\end{figure}

We present the \textbf{dispersion relations} obtained with a non vanishing Cosserat couple modulus $\mc>0$ (Figure \ref{DivNon}) and for a vanishing Cosserat couple modulus $\mc=0$ (Figure \ref{DivVan}). In the figures we consider  uncoupled waves (a), longitudinal waves (b) and
transverse waves (c). TRO: transverse rotational optic, TSO: transverse shear optic, TCVO:
transverse constant-volume optic, LA: longitudinal acoustic, LO$_{1}$-LO$_{2}$:
$1^{st}$ and $2^{nd}$ longitudinal optic, TA: transverse acoustic, TO$_{1}$-TO$_{2}$: $1^{st}$ and $2^{nd}$ transverse optic.

We can conclude that, when considering the micromorphic model with only $\rVert\Div\p\lVert^{2}$ for every value of   $\mu_{c}$, there always exist waves which propagate inside the considered medium independently of the value of the frequency. The uncoupled waves assume a peculiar behavior in which the frequency is independent of the wavenumber k.

\subsection{The relaxed micromorphic model obtained obtained  as a special case of the augmented relaxed model with $L_{d}=0$}

The relaxed micromorphic model is obtained by the model with $\rVert\Curl\p\lVert^{2}$ and $\rVert\Div\p\lVert^{2}$ by considering $L_{d}=0$ obtaining the energy:
\begin{align}
W=&\underbrace{\me\,\lVert \sym\left(\nablau-\p\right)\rVert ^{2}+\frac{\lle}{2}\left(\mathrm{tr} \left(\nablau-\p\right)\right)^{2}}_{\mathrm{{\textstyle isotropic\ elastic-energy}}}	+\hspace{-0.1cm}\underbrace{\mc\,\lVert \skew\left(\nablau-\p\right)\rVert ^{2}}_{\mathrm{\textstyle rotational\   elastic\ coupling}		}\hspace{-0.1cm} \label{eq:Ener-2}\\
& \quad 
+\underbrace{\mh\,\lVert \sym \p\rVert ^{2}+\frac{\lh}{2}\,\left(\mathrm{tr} \p\right)^{2}}_{\mathrm{{\textstyle micro-self-energy}}}
+\hspace{-0.2cm}\underbrace{\frac{\mLc}{2} \,\lVert \Curl \p\rVert^2}_{\mathrm{\textstyle isotropic\ curvature}}\,.
\nonumber  
\end{align}
The dynamical equilibrium equations are, see also \cite{madeo2015wave}:
\begin{align}
\rho\,  u_{,tt}=&\,\Div\,\sigma=\,\Div\left[2\,\me\, \sym\left(\nablau-\p\right)+2\,\mc \,\skew\left(\nablau-\p\right)+\lle\tr\left(\nablau-\p\right)\mathds{1}\right] , \nonumber \\
\eta  \p_{,tt}=&\,2\, \me \,\sym\left(\nablau-\p\right) +2 \,\mc\, \skew\left(\nablau-\p\right)+\lle\tr\left(\nablau-\p\right)\mathds{1}\label{eq:Dyn}\\&\ -\left[2\mh \X\sym \p +\lh \tr(\p)\mathds{1}\right]-\mLc  \Curl  \Curl \p.\nonumber 
\end{align}

\begin{figure}[H]
	\begin{centering}
		\begin{tabular}{ccccc}
			(a)\includegraphics[width=4cm]{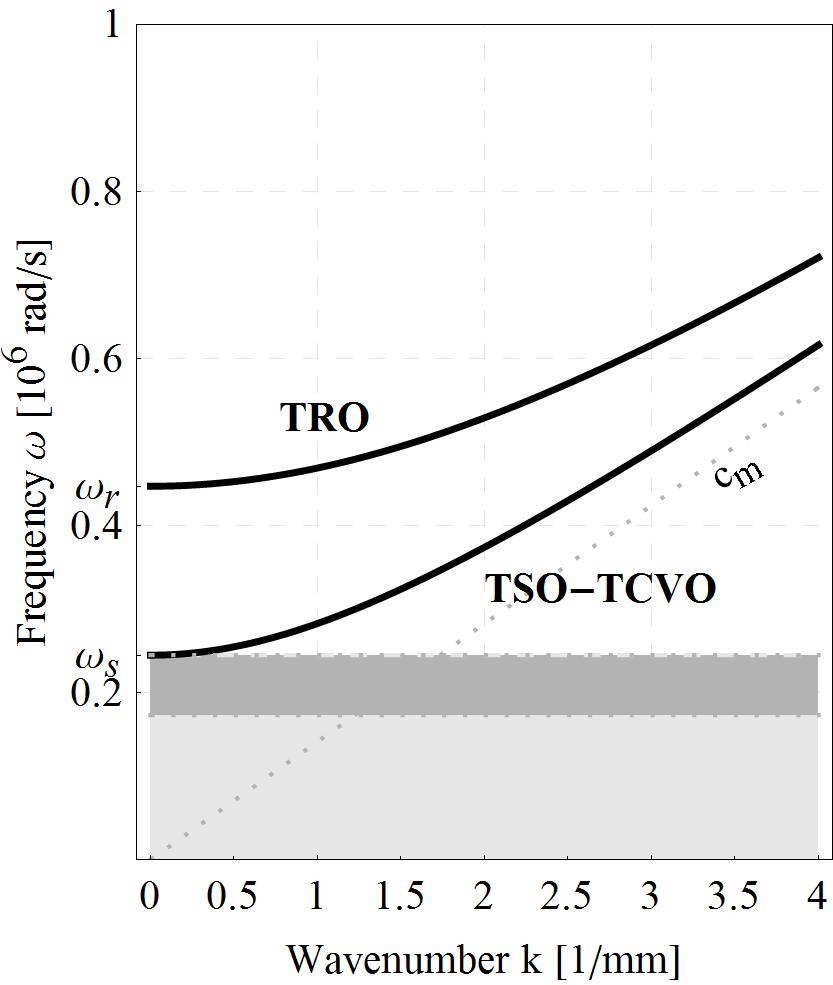}  &\qquad(b) \includegraphics[width=4cm]{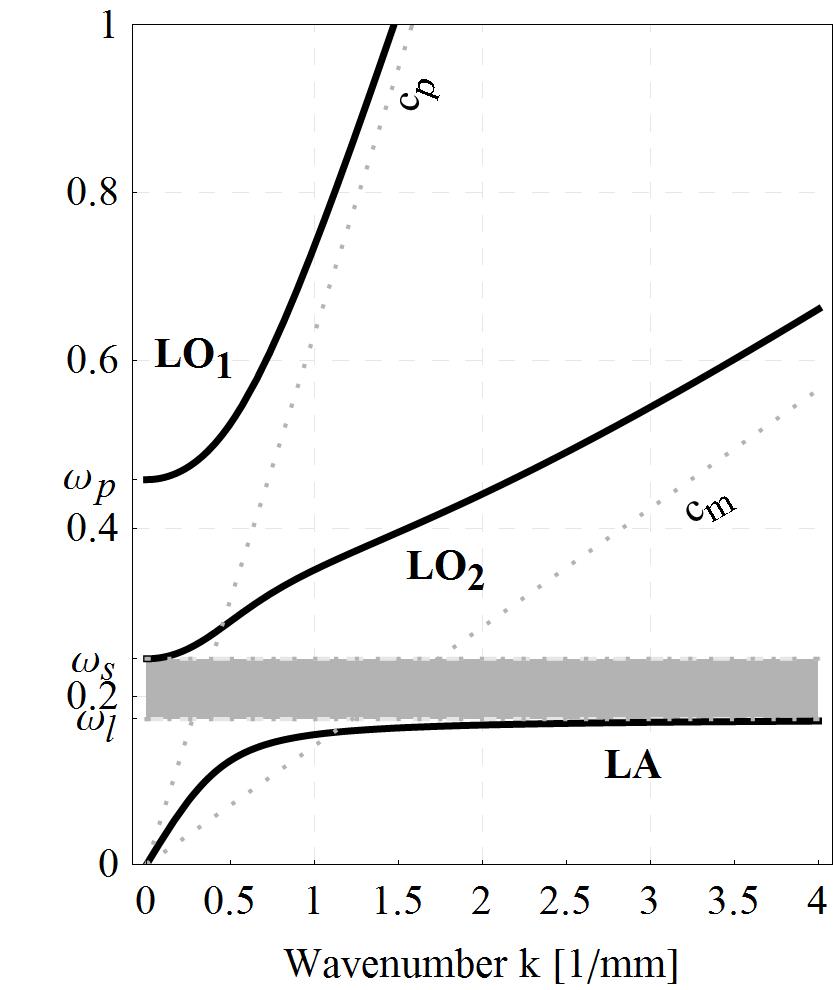} &\qquad(c) \includegraphics[width=4cm]{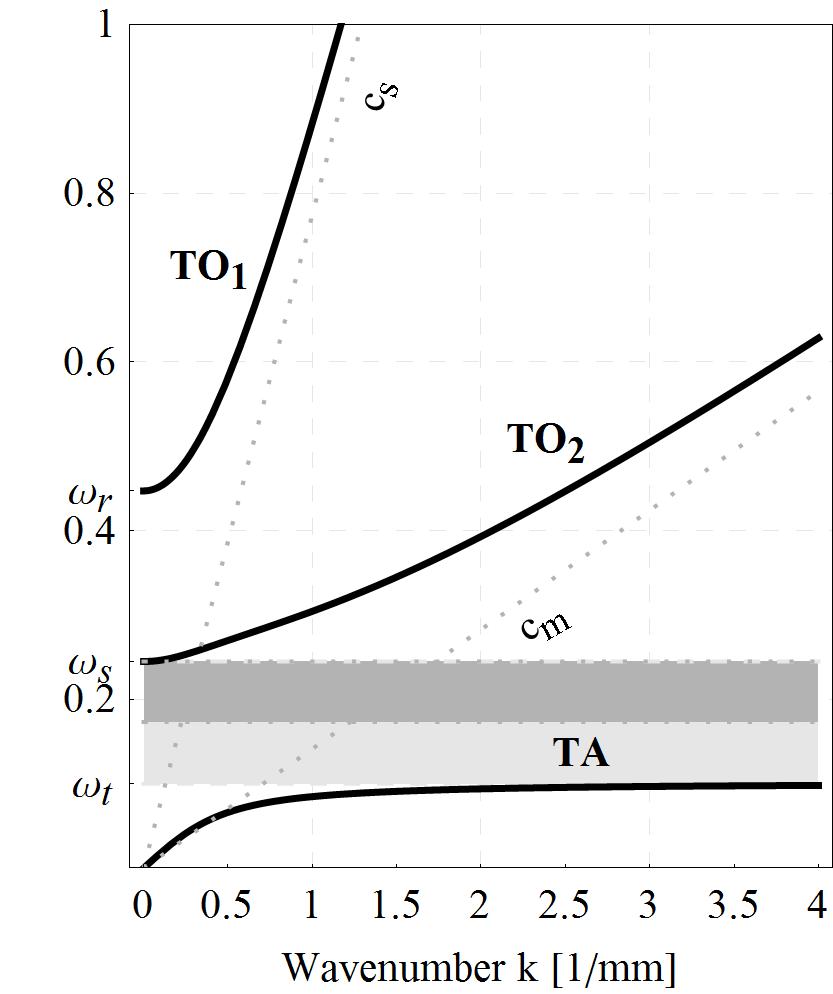} &
		\end{tabular}
		\par\end{centering}
	
	\caption{\label{CurlNon}Dispersion relations $\omega=\omega(k)$ for the
		\textbf{relaxed micromorphic model}  with non-vanishing Cosserat couple modulus $\mc>0$. \textbf{Complete frequency band gap} is the shaded intersected domain bounded from the maximum between $\omega_{l}$ and $\omega_{t}$ and the minimum between $\omega_{r}$ and $\omega_{s}$. The existence of the band gap is related to $\mc>0$ via the cut-off frequency $\omega_{r}=\sqrt{\frac{2\mc}{\eta}}$ of the uncoupled waves TRO and TO1   
	}	
\end{figure}

\begin{figure}[H]
	\begin{centering}
		\begin{tabular}{ccccc}
			(a)\includegraphics[width=4cm]{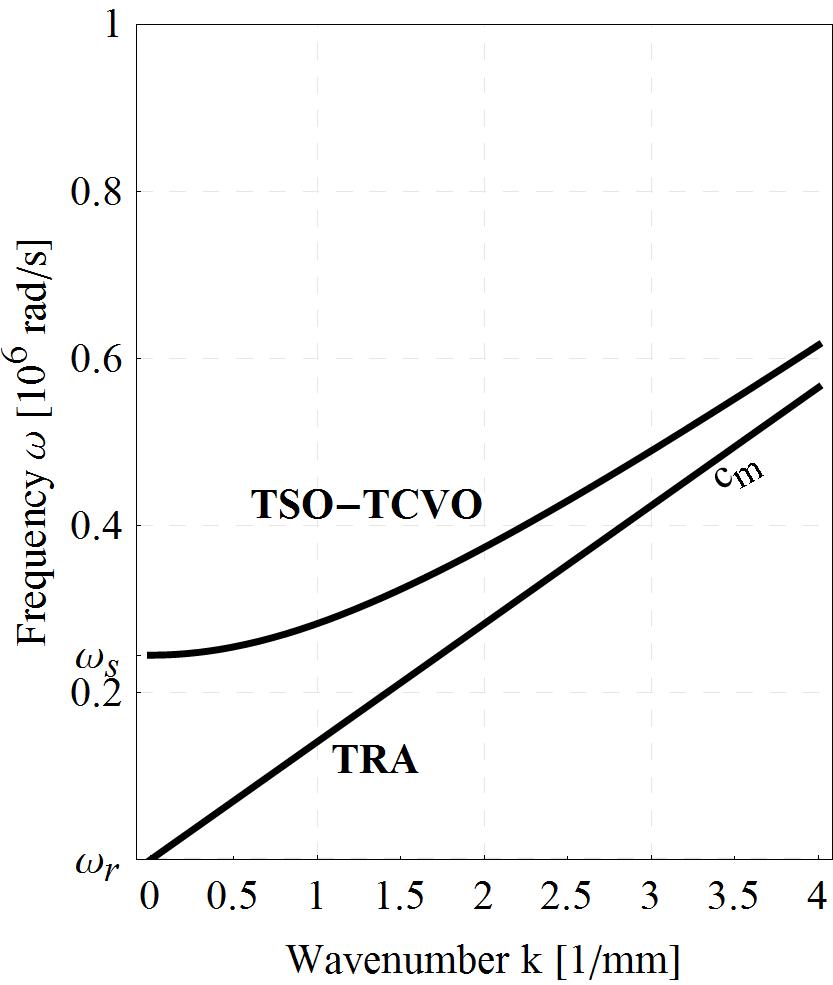}  &\qquad(b) \includegraphics[width=4cm]{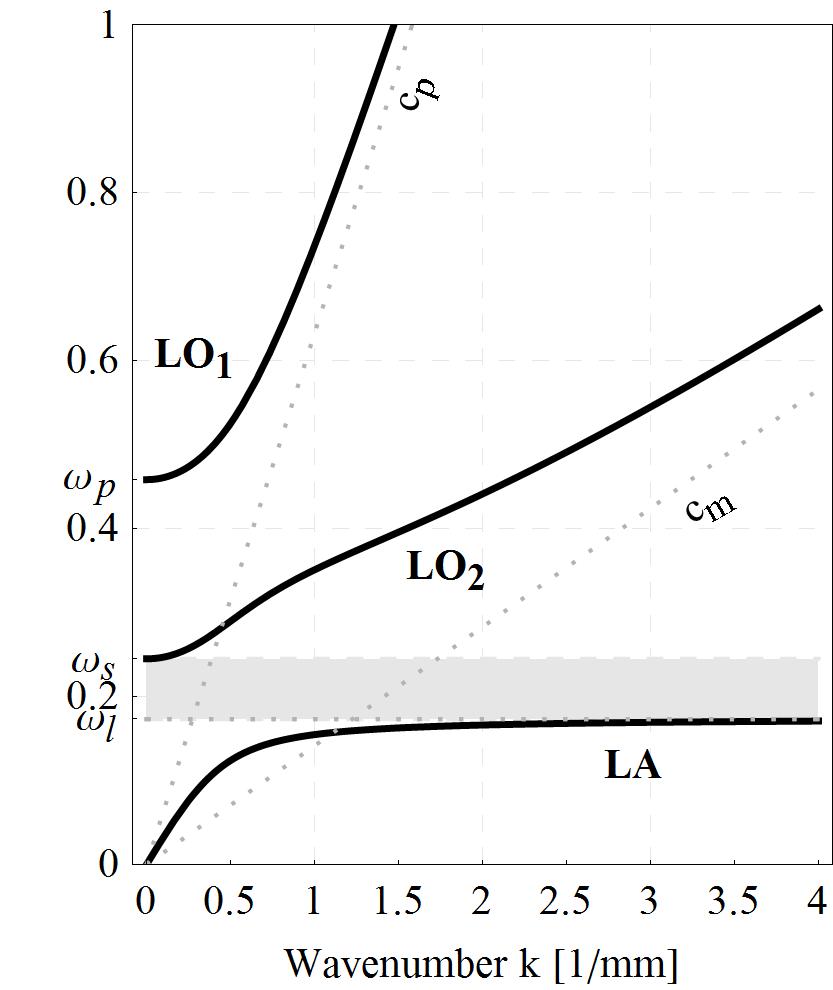} &\qquad(c) \includegraphics[width=4cm]{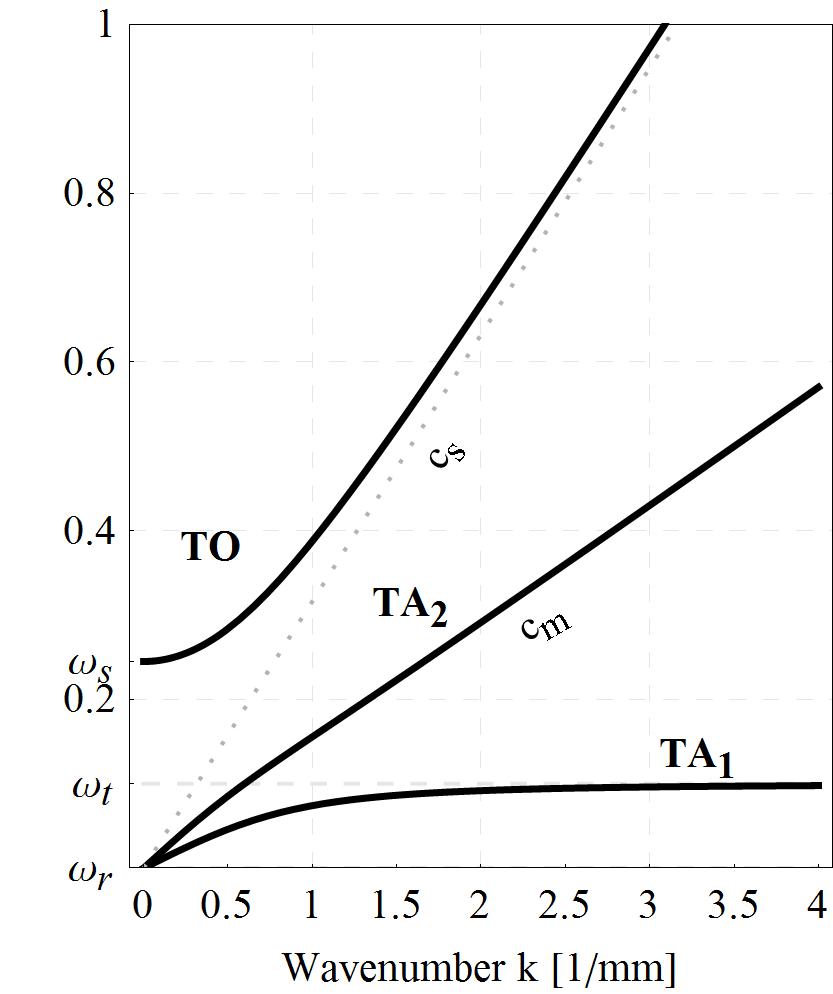} &
		\end{tabular}
		\par\end{centering}
	
	\caption{\label{CurlVan}Dispersion relations $\omega=\omega(k)$ for the
		\textbf{relaxed micromorphic model}  with vanishing Cosserat couple modulus $\mc=0$: only a \textbf{partial band gap} can be modeled.
	}	
\end{figure}

We present the \textbf{dispersion relations} obtained with a non vanishing Cosserat couple modulus $\mc>0$ (Figure \ref{CurlNon}) and for a vanishing Cosserat couple modulus $\mc=0$ (Figure \ref{CurlVan}). In the figures we consider  uncoupled waves (a), longitudinal waves (b) and
transverse waves (c). TRO: transverse rotational optic, TSO: transverse shear optic, TCVO:
transverse constant-volume optic, LA: longitudinal acoustic, LO$_{1}$-LO$_{2}$:
$1^{st}$ and $2^{nd}$ longitudinal optic, TA: transverse acoustic, TO$_{1}$-TO$_{2}$: $1^{st}$ and $2^{nd}$ transverse optic.

We can conclude that, in general, when considering the relaxed micromorphic medium with vanishing Cosserat couple modulus $\mu_{c}$, there always exist waves which propagate inside the considered medium independently of the value of the frequency. Nevertheless, if one considers a particular case (obtained by imposing suitable kinematical constraints) in which only longitudinal waves can propagate, then in the frequency range $\left(\omega_{s},\omega_{l}\right)$ only standing wave exist which do not allow for wave propagation.

On the other hand, switching on the Cosserat couple modulus $\mu_{c}$, allows for the description of complete frequency band-gaps in which no propagation can occur. 

\section{The standard Mindlin-Eringen model with $\rVert\nablap\lVert^{2}$\label{sec:Mind}}
The elastic energy of the general anisotropic centro-symmetric micromorphic model in
the sense of Mindlin-Eringen (see \cite{mindlin1964micro} and \cite[p. 270, eq. 7.1.4]{eringen1999microcontinuum}) can be represented
as:
\begin{align}
W= & \underbrace{\frac{1}{2}\langlenew\Coe \X \left(\nablau-\p\right),\left(\nablau-\p\right)\ranglenew_{\R^{3\times 3}}}_{\mathrm{{\textstyle full\ anisotropic\ elastic-energy}}}+\underbrace{\frac{1}{2}\langlenew\Ch \X\sym \p,\sym \p\ranglenew_{\R^{3\times 3}}}_{\mathrm{\textstyle micro-self-energy}}\label{eq:EnerEringen}\\
&+\underbrace{\frac{1}{2} \langlenew \E \X \left(\nablau- \p\right), \sym \p\ranglenew_{\R^{3\times 3}}}_{\mathrm{\textstyle anisotropic\ cross-coupling}}+\underbrace{\frac{\mLc}{2} \langlenew \Ls_{\text{aniso}} \X \nabla\hspace{-0.1cm}  \p,\nabla\hspace{-0.1cm} \p\ranglenew_{\R^{3\times 3\times 3}}}_{\mathrm{\textstyle full\ anisotropic\ curvature}}\,,\nonumber 
\end{align}
where $\Coe:\R^{3\times3}\rightarrow\R^{3\times3}$
is a $4^{th}$ order micromorphic elasticity tensor which has at most 45 independent coefficients and which acts on the \textbf{non-symmetric elastic distortion} $e=\nablau-\p$ and $\E:\R^{3\times3}\rightarrow\Sym(3)$ is a $4^{th}$ order cross-coupling tensor with the symmetry $\left(\E\right)_{ijkl}=\left(\E\right)_{jikl}$ having at most 54 independent coefficients. The fourth order tensor $\Ch:\Sym(3)\rightarrow\Sym(3)$
has the classical 21 independent coefficients of classical elasticity, while $\Ls_{	\text{aniso}}:\R^{3\times 3\times 3}\rightarrow\R^{3\times 3\times 3}$ is a $6^{th}$ order tensor that shows an astonishing 378 parameters. The parameter $\mu>0$ is a typical shear modulus and $L_{c}>0$ is one characteristic length, while $\Ls_{\text{aniso}}$ is, accordingly, dimensionless.

One of the major obstacles in using the micromorphic approach for specific materials is the impossibility to determine such multitude of new material coefficients. Not only is the huge number a technical problem, but also the interpretation of coefficients is problematic \cite{chen2003determining,chen2003connecting,chen2004atomistic}. Some of these coefficients are size-dependent while others are not. A purely formal approach, as it is often done, cannot be the final answer.

In what follows, we will consider a simplified isotropic energy:
\begin{align}
W=&\underbrace{\me\,\lVert \sym\left(\nablau-\p\right)\rVert ^{2}+\frac{\lle}{2}\left(\mathrm{tr} \left(\nablau-\p\right)\right)^{2}}_{\mathrm{{\textstyle isotropic\ elastic-energy}}}	+\hspace{-0.1cm}\underbrace{\mc\,\lVert \skew\left(\nablau-\p\right)\rVert ^{2}}_{\mathrm{\textstyle rotational\   elastic\ coupling}		}\hspace{-0.1cm} \label{eq:Ener-Tot}\\
& \quad 
+\underbrace{\mh\,\lVert \sym \p\rVert ^{2}+\frac{\lh}{2}\,\left(\mathrm{tr} \p\right)^{2}}_{\mathrm{{\textstyle micro-self-energy}}}
+\hspace{-0.4cm}\underbrace{\frac{\mLc}{2} \ \lVert \nablap\rVert^2\,.}_{\mathrm{\textstyle isotropic\ curvature}}
\nonumber  
\end{align}
The dynamical equilibrium equations are:
\begin{align}
\rho\,  u_{,tt}=&\,\Div\,\sigma=\,\Div\left[2\,\me\, \sym\left(\nablau-\p\right)+2\,\mc \,\skew\left(\nablau-\p\right)+\lle\tr\left(\nablau-\p\right)\mathds{1}\right] , \nonumber \\\label{eq:DynMind}
\eta  \p_{,tt}=&\,2\, \me \,\sym\left(\nablau-\p\right) +2 \,\mc\, \skew\left(\nablau-\p\right)+\lle\tr\left(\nablau-\p\right)\mathds{1}\\&\ -\left[2\mh \X\sym \p +\lh \tr(\p)\mathds{1}\right]+\mLc \, \underbrace{ \Div\, \nabla  \p }_{ \Delta\p}.\nonumber 
\end{align}

\begin{figure}[H]
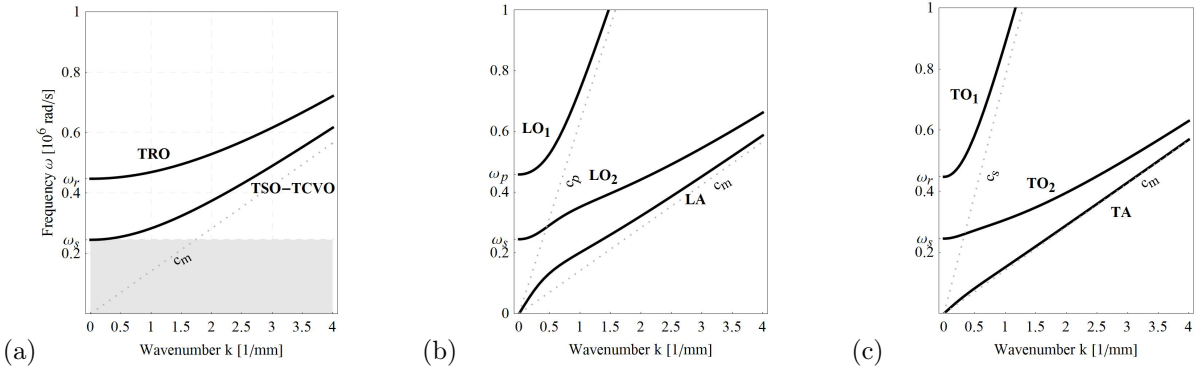

	\begin{centering}
		\begin{tabular}{ccccc}
			(a)\includegraphics[width=4cm]{CURLDIVUncoupled.jpg}  &\qquad(b) \includegraphics[width=4cm]{CURLDIVLongitudinal.jpg} &\qquad(c) \includegraphics[width=4cm]{CURLDIVTransverse.jpg} &
		\end{tabular}
		\par\end{centering}
	
	\caption{\label{MindNon}Dispersion relations $\omega=\omega(k)$ for the
		\textbf{standard micromorphic model with $\rVert\nablap\lVert^{2}$}  with non-vanishing Cosserat couple modulus $\mc>0$:  only a 	\textbf{partial band gap} can be modeled for uncoupled waves.
	}	
\end{figure}

\begin{figure}[H]
	\begin{centering}
		\begin{tabular}{ccccc}
			(a)\includegraphics[width=4cm]{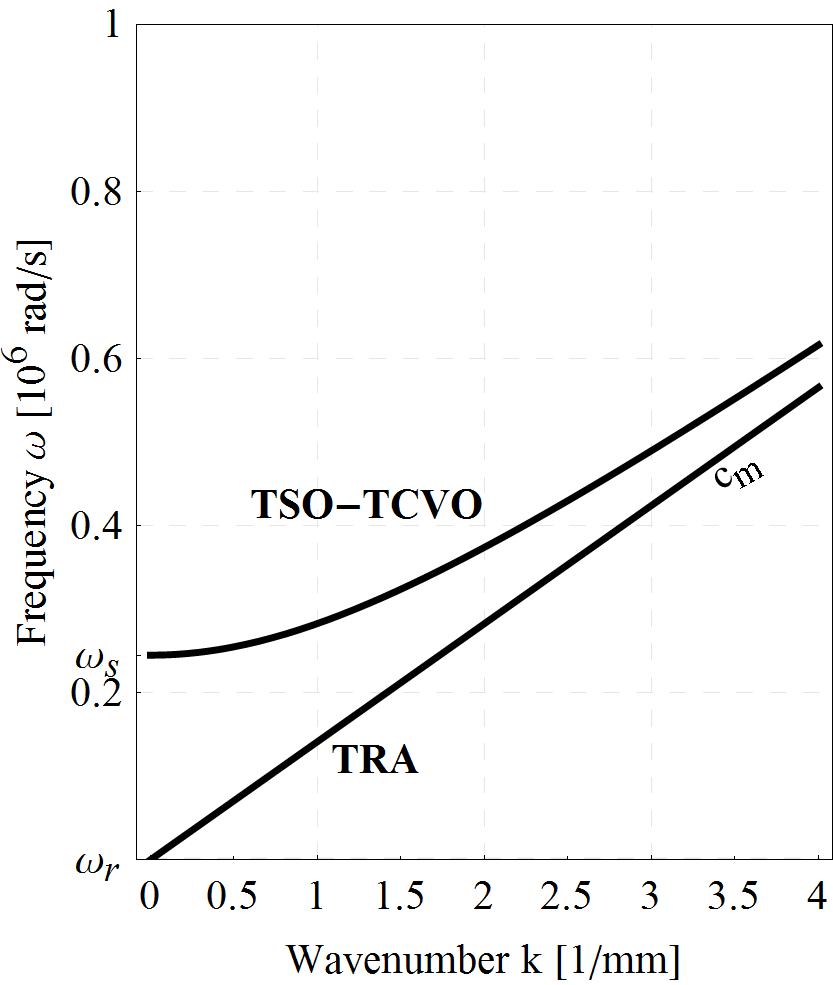}  &\qquad(b) \includegraphics[width=4cm]{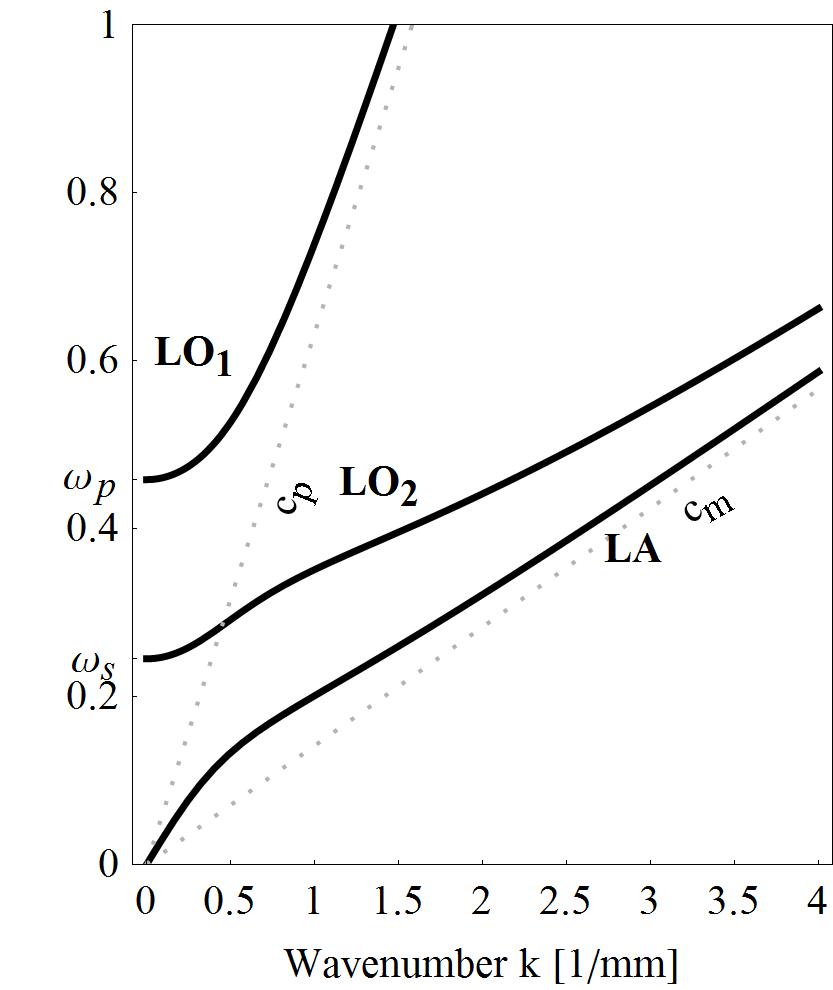} &\qquad(c) \includegraphics[width=4cm]{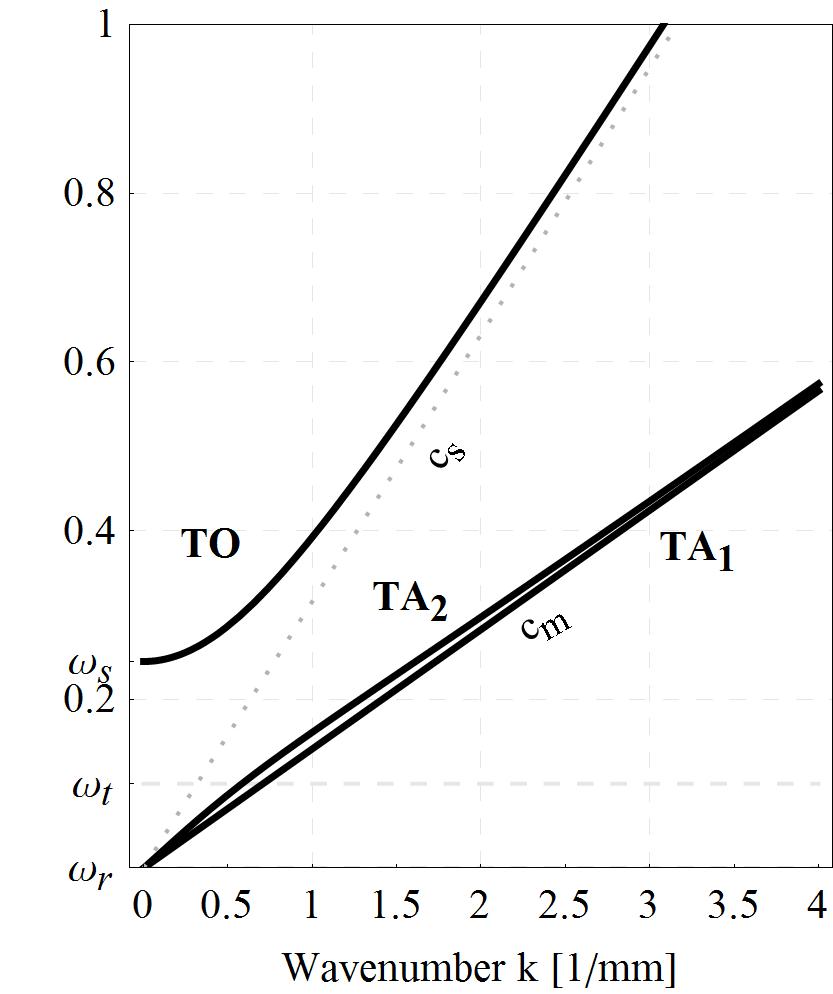} &
		\end{tabular}
		\par\end{centering}
	
	\caption{\label{MindVan}Dispersion relations $\omega=\omega(k)$ for the
		\textbf{standard micromorphic model with $\rVert\nablap\lVert^{2}$}  and vanishing Cosserat couple modulus $\mc=0$:  \textbf{no band gap} at all.
	}	
\end{figure}

We present the \textbf{dispersion relations} obtained with a non vanishing Cosserat couple modulus $\mc>0$ (Figure \ref{MindNon}) and for a vanishing Cosserat couple modulus $\mc=0$ (Figure \ref{MindVan}).  In the figures we consider  uncoupled waves (a), longitudinal waves (b) and
transverse waves (c). TRO: transverse rotational optic, TSO: transverse shear optic, TCVO:
transverse constant-volume optic, LA: longitudinal acoustic, LO$_{1}$-LO$_{2}$:
$1^{st}$ and $2^{nd}$ longitudinal optic, TA: transverse acoustic, TO$_{1}$-TO$_{2}$: $1^{st}$ and $2^{nd}$ transverse optic.

In a way completely equivalent to the case of $\rVert\Div\p\lVert^{2}$ and  $\rVert\Curl\p\lVert^{2}$ (see section \ref{DIVCURL}), we can conclude that when considering the standard Mindlin-Eringen model with vanishing Cosserat couple modulus $\mu_{c}$, there always exist waves which propagate inside the considered medium independently of the value of frequency even if considering separately  longitudinal, transverse and uncoupled waves.

The only effect obtainable switching on the Cosserat couple modulus $\mu_{c}$ is to obtain a partial band gap for the uncoupled waves.

\section{Conclusion}

\textbf{Metamaterials} are artifacts composed by \textbf{microstructural elements} assembled in periodic or quasi-periodic patterns, giving rise to materials with \textbf{unorthodox properties}. For some of these metamaterials, the presence of a microstructure allows for \textbf{local resonances} at the micro-level which globally result in \textbf{macroscopic wave-inhibition}: the energy of the incident wave remains trapped at the level of the microstructure. 

The presence of band gaps can be observed even in natural materials such as \textbf{perovskites}. 
Indeed, these materials are characterized by \textbf{microscopic rotational} and \textbf{stretch} motions  which can be observed using e.g. \textbf{Raman Spectroscopy}. The respective micro-vibrational modes, with frequencies  much higher than the acoustic modes,  give rise to some local resonances and thus to the onset of \textbf{band gaps}.

The \textbf{relaxed micromorphic model} is the \textbf{only linear, isotropic, reversibly elastic, non-local generalized continuum model} known to date able to \textbf{predict complete frequency band gaps}. It is decisive to use $\Curl \p$ instead of the full micro-distortion gradient $\nablap$ and to take a positive Cosserat couple modulus $\mc>0$. A material not showing band gaps should be modeled with $\mc\equiv0$.

Considering that non-locality is an intrinsic characteristic feature of micro-structured materials, especially when high contrasts of the mechanical properties occur at the micro-level, models that allow for its description are a necessary requirement. The relaxed micromorphic model is the only generalized continuum model which is simultaneously able to account for non-locality and for band-gaps onset in metamaterials.

\section{Acknowledgement}
The authors thank Samuel Forest (Centre des Matériaux, MINES Paristech) for drawing their attention to the question whether the "Curl" plays the decisive role in modeling band gaps.

Angela Madeo thanks INSA-Lyon for the funding of the BQR 2016 "Caractérisation mécanique inverse des métamatériaux: modélisation, identification expérimentale des paramètres et évolutions possibles".

\footnotesize

\let\stdsection\section
\def\section*#1{\stdsection{#1}}

\bibliography{library}
\bibliographystyle{plain}

\let\section\stdsection

\end{document}